\documentclass[authoryear,review,preprint,12pt]{elsarticle}
\usepackage[english,french]{babel}    
\usepackage[T1]{fontenc} 
\usepackage[cyr]{aeguill}
\usepackage{epsfig}
\usepackage{graphicx} 
\usepackage{epstopdf}
\usepackage{lscape} 
\usepackage[utf8]{inputenc}
\usepackage{multirow}

\usepackage[nottoc, notlof, notlot]{tocbibind} 
\usepackage{a4wide}
\usepackage{url}
\usepackage{color}
\usepackage{natbib}

\usepackage{lipsum}
\usepackage{multicol}
\usepackage{nopageno}
\usepackage{amsmath}
\usepackage{array}
\usepackage{amsthm}
\usepackage{numprint}
\usepackage{float}
\usepackage{amsfonts}
\usepackage{lineno}

\newcommand{\cn}{\mathrm{cn}}

\newcommand{\sn}{\mathrm{sn}}
\newcommand{\sd}{\mathrm{sd}}
\newcommand{\ns}{\mathrm{ns}}
\newcommand{\dn}{\mathrm{dn}}

\newcommand{\ds}{\mathrm{ds}}
\newcommand{\cs}{\mathrm{cs}}
\newcommand{\arcsn}{\mathrm{arcsn}}
\newcommand{\arccn}{\mathrm{arccn}}
\newcommand{\arccs}{\mathrm{arccs}}
\newcommand{\am}{\mathrm{am}}
\usepackage{numcompress}
\usepackage{fullpage}
\usepackage{empheq}
\usepackage{amssymb}
\usepackage{remreset}
\makeatletter \@removefromreset{footnote}{chapter} \makeatother
\usepackage {geometry}
\usepackage{subfigure}
\usepackage{listingsutf8}

\usepackage{wrapfig}
\begin{document}
\selectlanguage{english}

\begin{frontmatter}
\title{Theory for planetary exospheres: I. Radiation pressure effect on dynamical trajectories}
\author[ups,irap]{A.~Beth\corref{cor1}}
\ead{arnaud.beth@gmail.com}
\author[ups,irap]{P.~Garnier\corref{cor2}}
\ead{pgarnier@irap.omp.eu}
\author[ups,irap]{D.~Toublanc}
\author[ups,irap]{I.~Dandouras}
\author[ups,irap]{C.~Mazelle}
\address[ups]{Université de Toulouse; UPS-OMP; IRAP; Toulouse, France}
\address[irap]{CNRS; IRAP; 9 Av. colonel Roche, BP 44346, F-31028 Toulouse cedex 4, France}

\cortext[cor1]{Principal corresponding author}
\cortext[cor2]{Corresponding author}
\begin{abstract}
The planetary exospheres are poorly known in their outer parts, since the neutral densities are low compared with the instruments detection capabilities. The exospheric models are thus often the main source of information at such high altitudes. We present a new way to take into account analytically the additional effect of the radiation pressure on planetary exospheres. In a series of papers, we present with an Hamiltonian approach the effect of the radiation pressure on dynamical trajectories, density profiles and escaping thermal flux. Our work is a generalization of the study by \citet{Bishop1989}. In this first paper, we present the complete exact solutions of particles trajectories, which are not conics, under the influence of the solar radiation pressure. This problem was recently partly solved by \citet{Lantoine2011} and completely by \citet{Biscani2014}. We give here the full set of solutions, including solutions not previously derived, as well as simpler formulations for previously known cases and comparisons with recent works. The solutions given may also be applied to the classical Stark problem \citep{Stark1914}: we thus provide here for the first time the complete set of solutions for this well-known effect in term of Jacobi elliptic functions. 
\end{abstract}

\end{frontmatter}
\section{Introduction}
The exosphere is the upper layer of any planetary atmosphere: it is a quasi-collisionless medium where the particle trajectories are more dominated by gravity than by collisions. 
 Above the exobase, the lower limit of the exosphere, the Knudsen number \citep{Ferziger1972} becomes large, collisions become scarce, the distribution function cannot be considered as maxwellian anymore and, gradually, the trajectories of particles are essentially determined by the gravitation and radiation pressure by the Sun. The trajectories of particles, subject to the gravitational force, are completely solved with the equations of motion, but it is not the case with the radiation pressure \citep{Bishop1989}.

In the absence of radiation pressure, we can distinguish three types of trajectories for the exospheric particles :

\begin{itemize}
\item the escaping particles come from the exobase and have a positive mechanical energy: they can escape from the gravitational influence of the planet with a velocity larger than the escape velocity. These particles are responsible for the Jeans' escape \citep{Jeans1916}. They can also be defined as crossing only once the exobase,
\item the ballistic particles also come from the exobase but have a negative mechanical energy, they are gravitationally bound to the planet. They reach a maximum altitude and fall down on the exobase if they do not undergo collisions. They cross the exobase twice,
\item the satellite particles never cross the exobase. They also have a negative mechanical energy but their periapsis is above the exobase: they orbit along an entire ellipse around the planet without crossing the exobase. The satellite particles result in their major part from ballistic particles undergoing few collisions mainly near the exobase (\citet{Beth2014}). Thus, they do not exist in a collisionless model of the exosphere.
\end{itemize}

The radiation pressure disturbs the conics (ellipses or hyperbolas) described by the particles under the influence of gravity. The resonant scattering of solar photons leads to a total momentum transfer from the photon to the atom or molecule \citep{Burns1979}. In the non-relativistic case, assuming an isotropic reemission of the solar photon, this one is absorbed in the Sun direction and scattered with the same probability in all directions. For a sufficient flux of photons in the absorption wavelength range, the reemission in average does not induce any momentum transfer from the atom/molecule to the photon. The momentum variation, each second, between before and after the scattering imparts a force, the radiation pressure.

\citet{Bishop1989} proposed to analyze its effect on the structure of planetary exospheres. In particular, they highlighted analytically the ``tail" phenomenon at Earth: the density for atomic Hydrogen, which is sensitive to the Lyman-$\alpha$ photons, is higher in the nightside direction than in the dayside direction in the Earth corona. Nevertheless, their work was limited only to the Sun-planet axis, with a null component assumed for the angular momentum around the Sun-planet axis. We thus generalize here their work to a full 3D calculation, in order to investigate the influence of the radiation pressure on the trajectories (this paper), as well as the density profiles and escape flux (following works).

This problem is similar to the so-called Stark effect \citep{Stark1914}: the effect of a constant electric field on the atomic Hydrogen's electron. Its study can be transposed to celestial mechanics in order to describe the orbits of artificial and natural satellites in the perturbed (e.g. by the radiation pressure force) Two-Body Problem. A recent description of the Stark effect solutions was already given by \citet{Lantoine2011}. However, they give the analytical solutions of all trajectories only in the 2D case (and for bounded trajectories in the 3D case), and their formulas have some issues as will be discussed later. Another analytical study proposed by \citet{Biscani2014} uses the Weierstrassian formulations to solve the motions for bounded and unbounded trajectories and to find periodic motions. Also, the motion can be approached numerically by developing the equations of motion in Taylor series but this leads to some issues for high eccentricities \citep{Pellegrini2014}. \citet{Hatten2014} compared recently these methods and their computing efficiencies.

In this paper, based on the same formalism as \citet{Bishop1989}, we provide for the first time the complete exact 3D solutions of the Stark effect (and its celestial mechanics analogue) for any initial condition and for both bounded and unbounded trajectories. 

The first section describes the formalism used, before the sections \ref{model}/\ref{section3}/\ref{Time} provide the equations of motion and time. We then discuss about circular orbits in section \ref{circular}, while a comparison with previous works is given in section 6, before we conclude in section \ref{summary}.

\section{Model}\label{model}

In this work, we decide to study the effect of the radiation pressure on atomic Hydrogen in particular. Nevertheless, this formalism can be applied to any species subject to this force or to the interplanetary dust. We model the radiation pressure by a constant acceleration $a$ coming from the Sun. According to \citet{Bishop1991}, this acceleration depends on the line center solar Lyman-$\alpha$ flux $f_0$, in $10^{11}$ photons.cm$^{-2}$.s$^{-1}$.{\AA}$^{-1}$:
\begin{equation}
a=0.1774\ f_0\ (\text{cm.s}^{-2})
\end{equation}

\begin{figure}[!h]
\centering
\includegraphics[height=.60\linewidth]{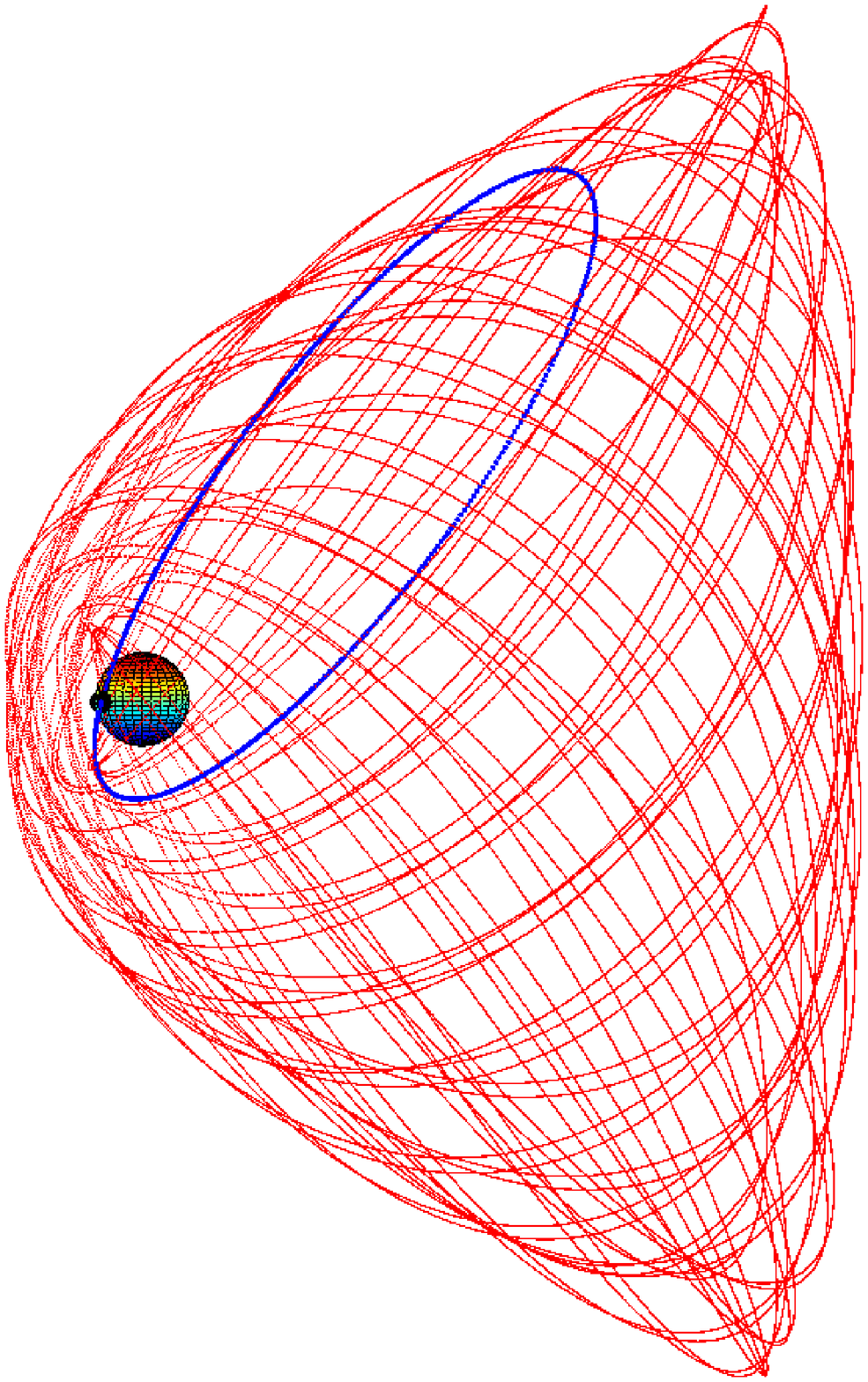}
\hfill
\includegraphics[height=.60\linewidth]{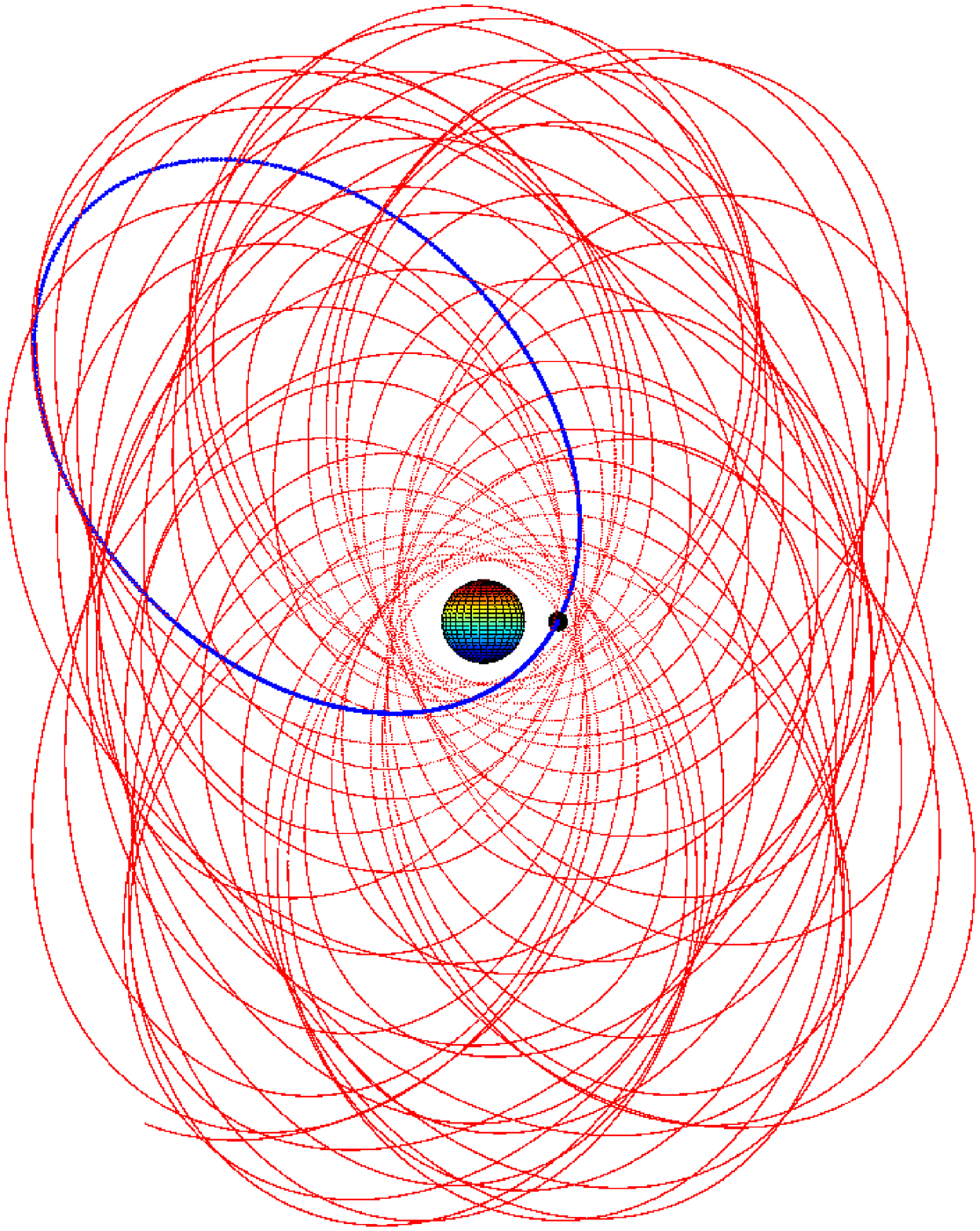}
\caption{Example of the trajectory of an atomic Hydrogen in the Earth exosphere. Left panel: view from the side. Right panel: view from the Sun.} 
\label{trajectoire}
\end{figure}

In spherical coordinates, the Hamiltonian of one Hydrogen atom can be written:
\begin{equation}
\begin{array}{l}
\mathcal{H}(r,\theta,\phi,p_r,p_{\theta},p_{\phi},t)\\\\
=\dfrac{p_{r}^2}{2m}+\dfrac{p_{\theta}^2}{2mr^2}+\dfrac{p_{\phi}^2}{2mr^2\sin^2 \theta}-\dfrac{GMm}{r}+mar\cos\theta
\label{hamiltonr}
\end{array}
\end{equation}
with $r$ the distance from the planet, $\theta$ the solar angle, $\phi$ the angle with respect to the ecliptic plane, $p_r$, $p_\theta$ and $p_\phi$ the conjugate momenta. $-GMm/r$ represents the gravitational potential and $mar\cos \theta$ the potential energy from the radiation pressure acceleration $a$. An example of trajectory of a $H$ atom subject to the radiation pressure is given in the figure \ref{trajectoire}.

This problem is similar to the classical Stark effect \citep{Stark1914}: a constant electric field (here the radiation pressure) is applied to an electron (here an Hydrogen atom) attached to a proton (here the planet). Both systems are equivalent because the force applied by the proton (the planet) to the electron (the Hydrogen atom), i.e. the electrostatic force, varies in $r^{-2}$ as the gravitational force from the planet on the Hydrogen atom. Thus, we adopt the same formalism as \citet{Sommerfeld1934} and use the parabolic coordinates. We use the transformation:
\begin{equation}
\begin{array}{rcccl}
u&=&r+x&=&r(1+\cos\theta)\\
w&=&r-x&=&r(1-\cos\theta)
\end{array}
\end{equation}
with $u$ and $w$ always positive.
Consequently, the Hamiltonian becomes:
\begin{equation}
\begin{array}{l}
\mathcal{H}(u,w,p_u,p_w,p_\phi)\\\\
=\dfrac{2up^{2}_u+2wp^{2}_{w}}{m(u+w)}+\dfrac{p^{2}_{\phi}}{2muw}-\dfrac{2GMm}{u+w}+ma\dfrac{u-w}{2}
\label{hamiltonuw}
\end{array}
\end{equation}
independent from $t$ and $\phi$.

According to hamiltonian canonical relations, we have:
\begin{equation}
\begin{array}{lcl}
\displaystyle{p_{u}}&=&\dfrac{m(u+w)}{4u}\dfrac{\mathrm{d} u}{\mathrm{d}t} \\\\
\displaystyle{p_{w}}&=&\dfrac{m(u+w)}{4w}\dfrac{\mathrm{d} w}{\mathrm{d}t} \\\\
\displaystyle{p_{\phi}}&=&m uw \dfrac{\mathrm{d}\phi}{\mathrm{d}t} 
\end{array}
\end{equation}

\subsection{Constants of the motion}
In this new system of coordinates, we study this Hamiltonian. First, $\mathcal{H}$ is independent from $t$ explicitly. $\mathcal{H}$ is a conserved quantity along the time and corresponds to the mechanical energy $E$ of the system. Moreover, as $\mathcal{H}$ is independent from $\phi$, according to canonical relations:
\begin{equation}
\dfrac{\mathrm{d} p_\phi}{\mathrm{d}t}=-\dfrac{\mathrm{d}\mathcal{H}}{\mathrm{d}\phi}=0
\end{equation}

Thus, $p_{\phi}$ is another constant of the motion.
Once $E$ and $p_\phi$ defined, the equation \ref{hamiltonuw} can be rewritten:
\begin{equation}
\begin{array}{l}
2muE-4up^{2}_u-\dfrac{p^{2}_{\phi}}{u}-m^2au^2+2GMm^2\\\\
=-2mwE+4wp^{2}_w+\dfrac{p^{2}_{\phi}}{w}-m^2aw^2-2GMm^2 
\end{array}
\end{equation}

The left hand side is a function dependent only on $u$ and $p_u$, the right hand side depends only on $w$ and $p_w$. As both functions are equal and independent, they are equal to a constant $A$, a separation constant:
\begin{equation}
\begin{array}{rcl}
A&=&2muE-4up^{2}_u-\dfrac{p^{2}_{\phi}}{u}-m^2au^2+2GMm^2\\\\
&=&-2mwE+4wp^{2}_w+\dfrac{p^{2}_{\phi}}{w}-m^2aw^2-2GMm^2 
\end{array}
\label{A}
\end{equation}

The motion possesses three constants: $E$, $A$ and $p_\phi$. The equation \ref{A} allows to express $p_u$ (respectively $p_w$) as a functions of $E$, $A$, $p_\phi$ and $u$ (respectively $w$):

\begin{equation}
\begin{array}{rcl}
p_u&=&\pm\sqrt{\dfrac{-P_3(u)}{4u^2}}\\\\
p_w&=&\pm\sqrt{\dfrac{Q_3(w)}{4w^2}}\\
\end{array}
\label{pupw}
\end{equation}
with
\begin{equation}
\begin{array}{rcl}
P_3(u)&=&mau^3-2mEu^2-(2GMm^2-A)u+p^{2}_{\phi}\\\\
Q_3(w)&=&maw^3+2mEw^2+(2GMm^2+A)w-p^{2}_{\phi}\\
\end{array}
\end{equation}

\subsection{Effective potentials}

We have already introduced the Hamiltonian $\mathcal{H}$ of the system. We can extend the approach according to Hamilton-Jacobi equations:
\begin{equation}
\begin{array}{rcl}
\dfrac{\partial \mathcal{S}}{\partial q_i}&=&p_i\\\\
\dfrac{\partial \mathcal{S}}{\partial t}&=&-\mathcal{H}
\end{array}
\end{equation}
where $\mathcal{S}$ is the Hamilton's principal function or action. This function depends on initial conditions (as $u_0$, $w_0$, $\phi_0$ and $t_0$) and the actual position of the particle (as $u$, $w$, $\phi$ and $t$). As previously demonstrated, $\mathcal{H}=E$ and $p_{\phi}$ are constants. Thus,
\begin{equation}
\begin{array}{rcl}
\dfrac{\partial \mathcal{S}}{\partial \phi}&=&p_\phi\\\\
\dfrac{\partial \mathcal{S}}{\partial t}&=&-E
\end{array}
\end{equation}
and leads to:
\begin{equation}
\mathcal{S}=-E(t-t_0)+p_{\phi}(\phi-\phi_0)+\hat{\mathcal{S}}[u_0,w_0,u,w,E,p_{\phi}]
\end{equation}
with $\hat{\mathcal{S}}$ the part of the action independent from $t_0$, $t$, $\phi_0$ and $\phi$.\newline

Moreover, the action $\hat{\mathcal{S}}$ can be separated into two parts: one written as a function of $u$ and $p_u$ coordinates, the other one with $w$ and $p_w$. By definition, according to the Hamilton-Jacobi equations, we have:
\begin{equation}
\begin{array}{rcccl}
\dfrac{\partial \mathcal{S}}{\partial u}&=&\dfrac{\partial \hat{\mathcal{S}}}{\partial u}&=&p_u\\\\
\dfrac{\partial \mathcal{S}}{\partial t}&=&\dfrac{\partial \hat{\mathcal{S}}}{\partial w}&=&p_w
\end{array}
\end{equation}
with $p_u$ (resp. $p_w$) a function only of $u$ (resp. $w$), assuming $E$, $A$ and $p_\phi$ values already fixed by initial conditions. Then, we can separate again the action, leading to:
\begin{equation}
\hat{\mathcal{S}}[u,w,u_0,w_0,E,p_{\phi}]=\mathcal{S}_{u}[u,E,A,p_{\phi}]+\mathcal{S}_{w}[w,E,A,p_{\phi}]
\end{equation}
\begin{equation}
\begin{array}{rcl}
\dfrac{\partial \mathcal{S}_u}{\partial u}&=&p_u\\\\
\dfrac{\partial \mathcal{S}_w}{\partial w}&=&p_w
\end{array}
\end{equation}
According to the equation \ref{pupw}, we have the following relations:
\begin{equation}
\begin{array}{lcl}
\displaystyle{\left( \frac{\mathrm{d}\mathcal{S}_u}{\mathrm{d} u}\right)^2}&=&\displaystyle{\frac{m}{2}\left(E-\frac{p_{\phi}^2}{2mu^2}+\frac{GMm}{u}-\frac{A}{2mu}-\frac{mau}{2}\right)}\\\\
&=&\displaystyle{\frac{m}{2}(E-V_u(u))}>0\\\\
\displaystyle{\left( \frac{\mathrm{d}\mathcal{S}_w}{\mathrm{d} w}\right)^2}&=&\displaystyle{\frac{m}{2}\left(E-\frac{p_{\phi}^2}{2mw^2}+\frac{GMm}{w}+\frac{A}{2mw}+\frac{maw}{2}\right)}\\\\
&=&\displaystyle{\frac{m}{2}(E-V_w(w))}>0
\label{potentiel}
\end{array}
\end{equation}
with
\begin{equation}
\begin{array}{lcl}
\displaystyle{V_u(u)}&=&\displaystyle{\frac{p_{\phi}^2}{2mu^2}-\frac{GMm}{u}+\frac{A}{2mu}+\frac{mau}{2}}\\\\
\displaystyle{V_w(w)}&=&\displaystyle{\frac{p_{\phi}^2}{2mw^2}-\frac{GMm}{w}-\frac{A}{2mw}-\frac{maw}{2}}
\end{array}
\label{VuVw}
\end{equation}
$V_u$ and $V_w$ are effective potentials applied in $u$ and $w$ directions (represented in the figure \ref{potentiel}). These potentials play key roles for the motion because they constrained the motion in $u$ and $w$ directions independently. For the motion of the particle, we must respect two conditions: $E>V_{u}(u)$ and $E>V_{w}(w)$. These conditions are analogous to:
\begin{equation}
\begin{array}{ccc}
P_3(u)<0&\text{and}&Q_3(w)>0
\end{array}
\label{conditions}
\end{equation}

\begin{figure}[!h]
\centering
\includegraphics[width=\linewidth]{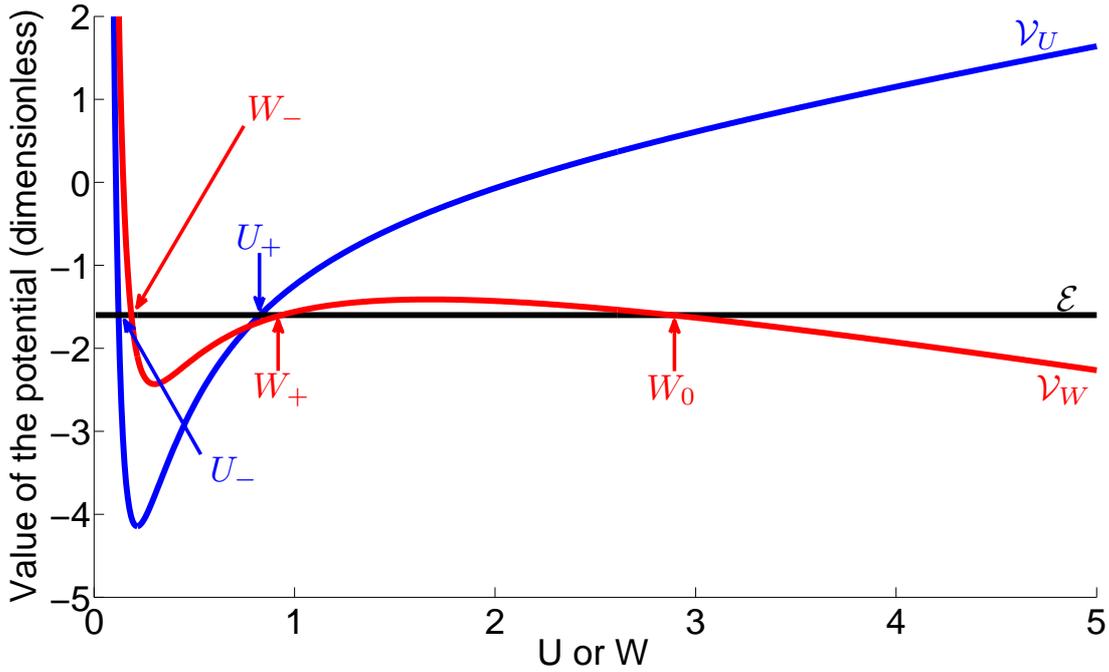}
\caption{Representation in dimensionless unity of the shape of both potentials $V_{U}$ (corresponding to $\mathcal{V}_U$, blue line) and $V_W$ (corresponding to $\mathcal{V}_W$, red line) for a set of $E$, $A$ and $p_{\phi}$ values. The motion is possible only in the area where the potential is below the mechanical energy $E$, represented by the black horizontal line. The different roots of $P_3$ and $Q_3$ are displayed and correspond to the intersection of the potentials with the horizontal black line. $U_{-}$ is the dimensionless value referring to $u_{-}$ (cf. section \ref{dimensionless_section}, table \ref{dimensionless}). Notice that $\mathcal{V}_W$ will cross only once the energy level $\mathcal{E}$ for too high or too low $\mathcal{E}$ values. }
\label{potentiel}
\end{figure}

These both conditions are more restrictive than the usual $E>E_p$ where $E_p$ is the potential energy.

\subsection{Study of $P_3$}

$P_3$ is a polynom of degree 3 with $\lim\limits_{u\to+\infty}P_{3}(u)=+\infty$. This polynom possesses three roots, whose one is real at least. As $P_3(0)=p_{\phi}^{2}>0$, one of these roots is real negative, according to intermediate value theorem since $\lim\limits_{u\to-\infty}P_{3}(u)=-\infty$. Nevertheless, the motion occurs for positive $u$ values, and we know this motion exists. It implies there is an interval in $\mathbb{R}^{+}$ such as $P_3<0$ (otherwise, there is no motion in $u$-direction, no possible physically, cf. eq. \ref{conditions}). To comply with this last condition, both other roots are real and positive. In summary, $P_3$ has three real roots: one negative and two positive.

We call each root $u_0$, $u_{-}$ and $u_{+}$ such as $u_0<0<u_{-}<u_{+}$ and the $u$-motion is restricted to $u\in[u_{-};u_{+}]$ ($U\in[U_{-};U_{+}]$ in term of dimensionless quantities, as can be seen in the figure \ref{potentiel}).

\subsection{Study of $Q_3$}

$Q_3$ is a polynom of degree 3 with $\lim\limits_{w\to+\infty}Q_{3}(w)=+\infty$. This polynom possesses three roots, whose one is real at least. As $Q_3(0)=-p_{\phi}^{2}<0$, one of these roots is real positive, according to intermediate value theorem since $\lim\limits_{w\to+\infty}Q_{3}(w)=+\infty$. Nevertheless, the motion occurs for positive $w$ values. We have restrictions on both other roots: they must be both real positive, both real negative or both complex conjugates.

In the case where the three roots are real positive, we call each root $w_0$, $w_{-}$ and $w_{+}$ such as $0<w_{-}<w_{+}<w_{0}$ and the motion is restricted to $w\in[w_{-};w_{+}]\cup[w_{0};+\infty[$ (as can be seen in the figure \ref{potentiel} with the dimensionless quantities, cf. section \ref{dimensionless_q}).

In the case with one positive root and both other complex or real negative, we call each root $w_0$ (the positive one), $w_{-}$ and $w_{+}$ (keep the same order as previously defined if they are real) such as the motion is restricted to $w\in[w_{0};+\infty[$ only ($[W_0;+\infty[$ in term of dimensionless quantities).

\subsection{Restriction on the motion}
\begin{figure}[!h]
\centering
\includegraphics[width=.4\linewidth]{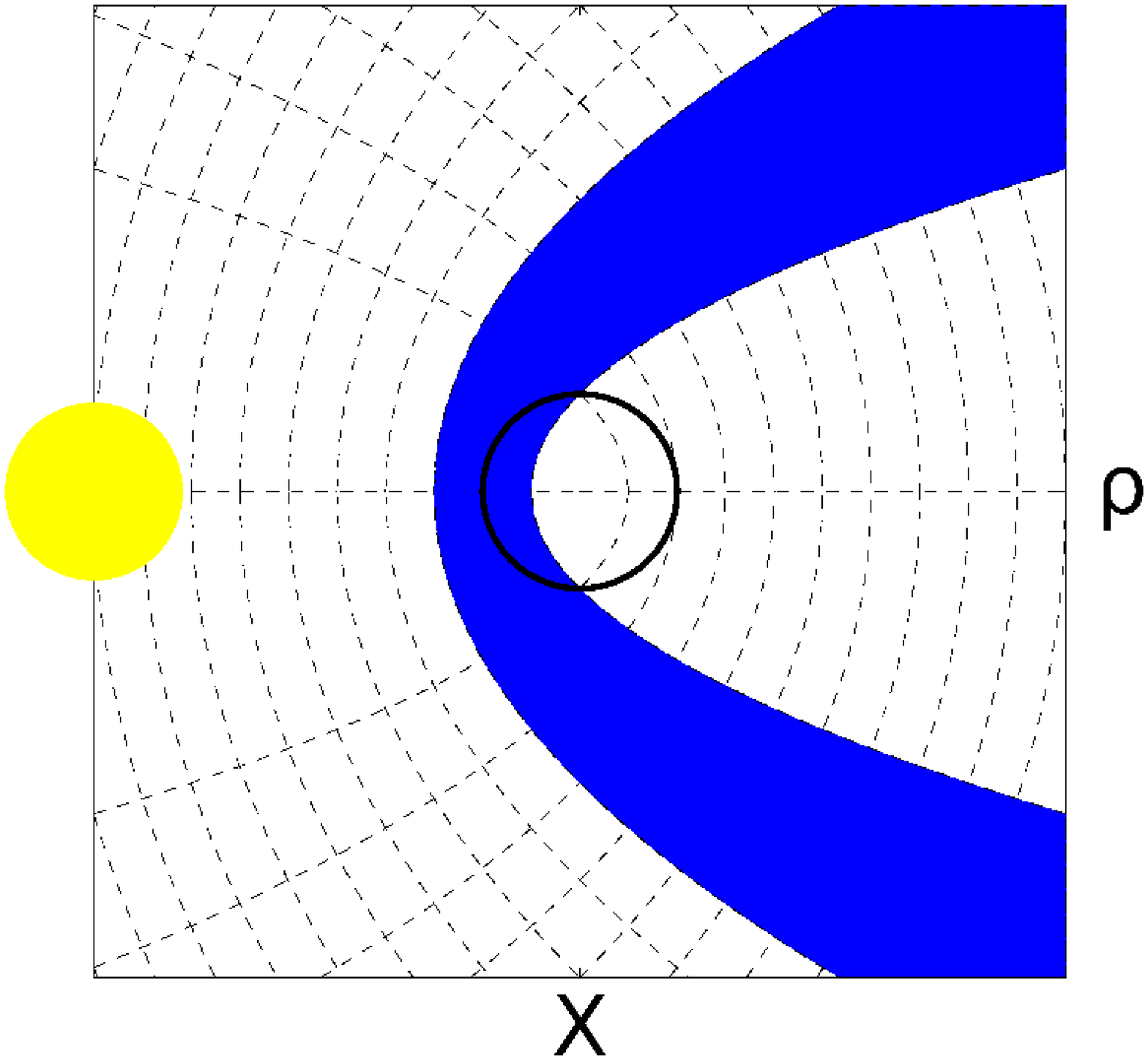}\includegraphics[width=.4\linewidth]{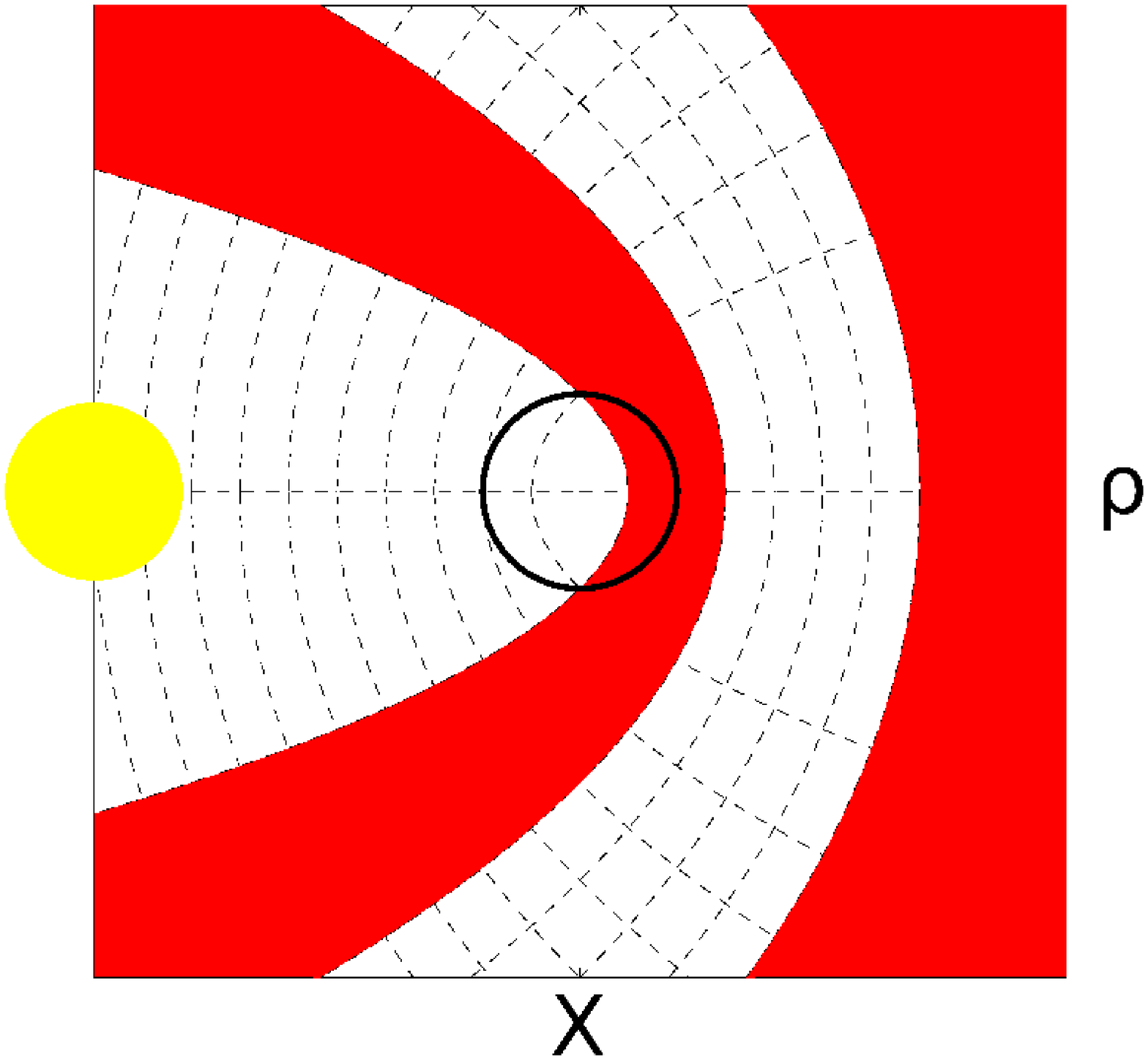}
\caption{Representation of the available space for the motion. Left panel: the blue area corresponds to the allowed region for the $u$-motion. Right panel: the red area corresponds to the allowed region for the w-motion. The regions are limited by paraboloids (parabolas here by projection). The Sun location is shown with a yellow circle.}
\label{restriction1}
\end{figure}

Each constant value of $u$ or $w$ defines a paraboloid in three dimensions. For each interval, constrained by fixed values of $u$ and $w$, the motion will be contained between the paraboloids defined by these limit values as shown in the figure \ref{restriction1}.

For the $u$-motion, this is always limited by two paraboloids defined by $u=u_{-}$ and $u=u_{+}$ as shown by the blue area in the figure \ref{restriction1} (left panel). Similarly, for the $w$-motion, there are two cases: the motion is constrained between one paraboloid ($w=w_{0}$) and the infinity or between two paraboloids ($w=w_{-}$ and $w=w_{+}$). Both cases are represented by the red area in the figure \ref{restriction1} (right panel).

\subsection{Summary of restrictions}\label{dimensionless_q}

\begin{figure}[!h]
\centering
\includegraphics[width=.4\linewidth]{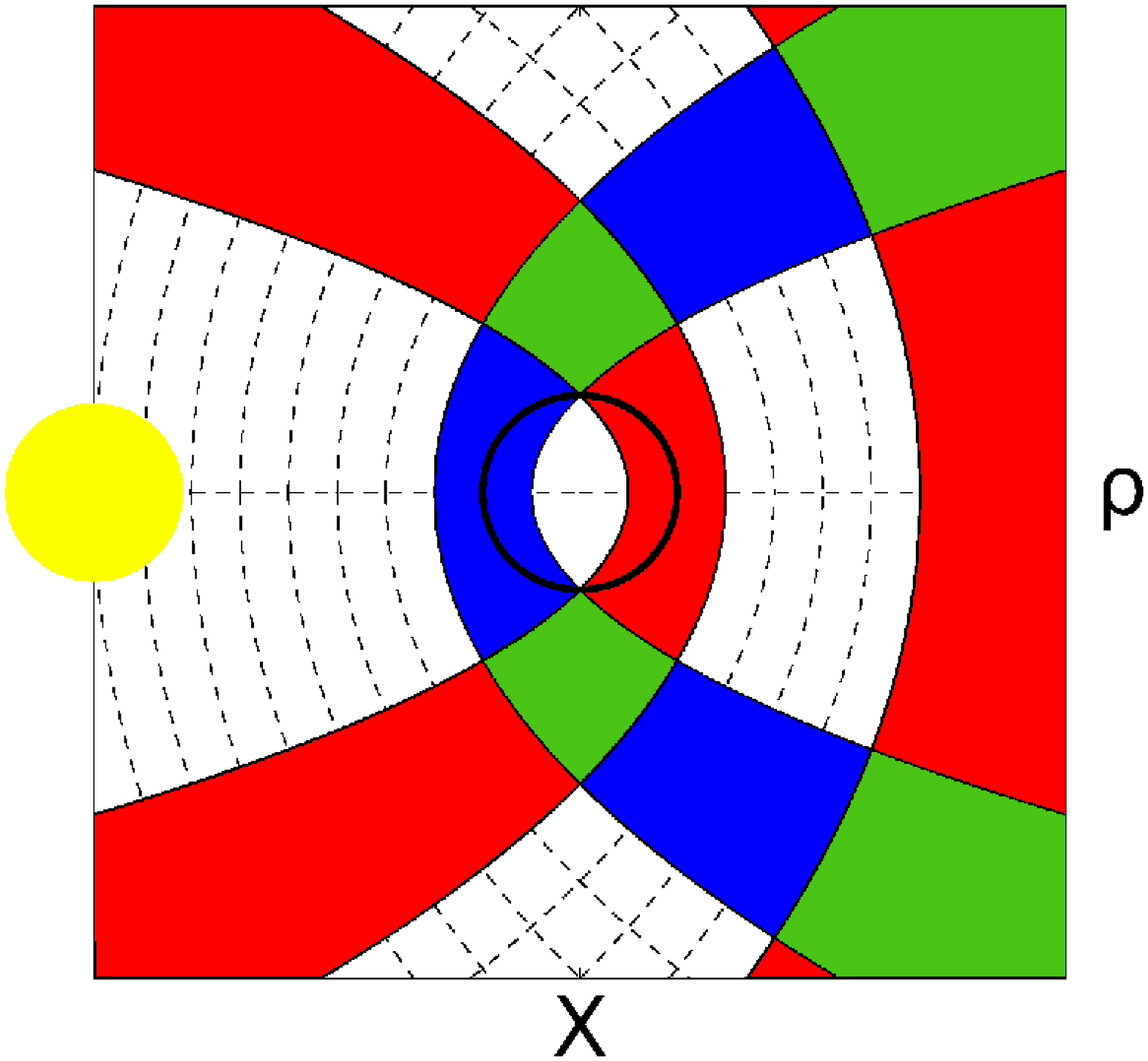}
\includegraphics[width=.43\linewidth]{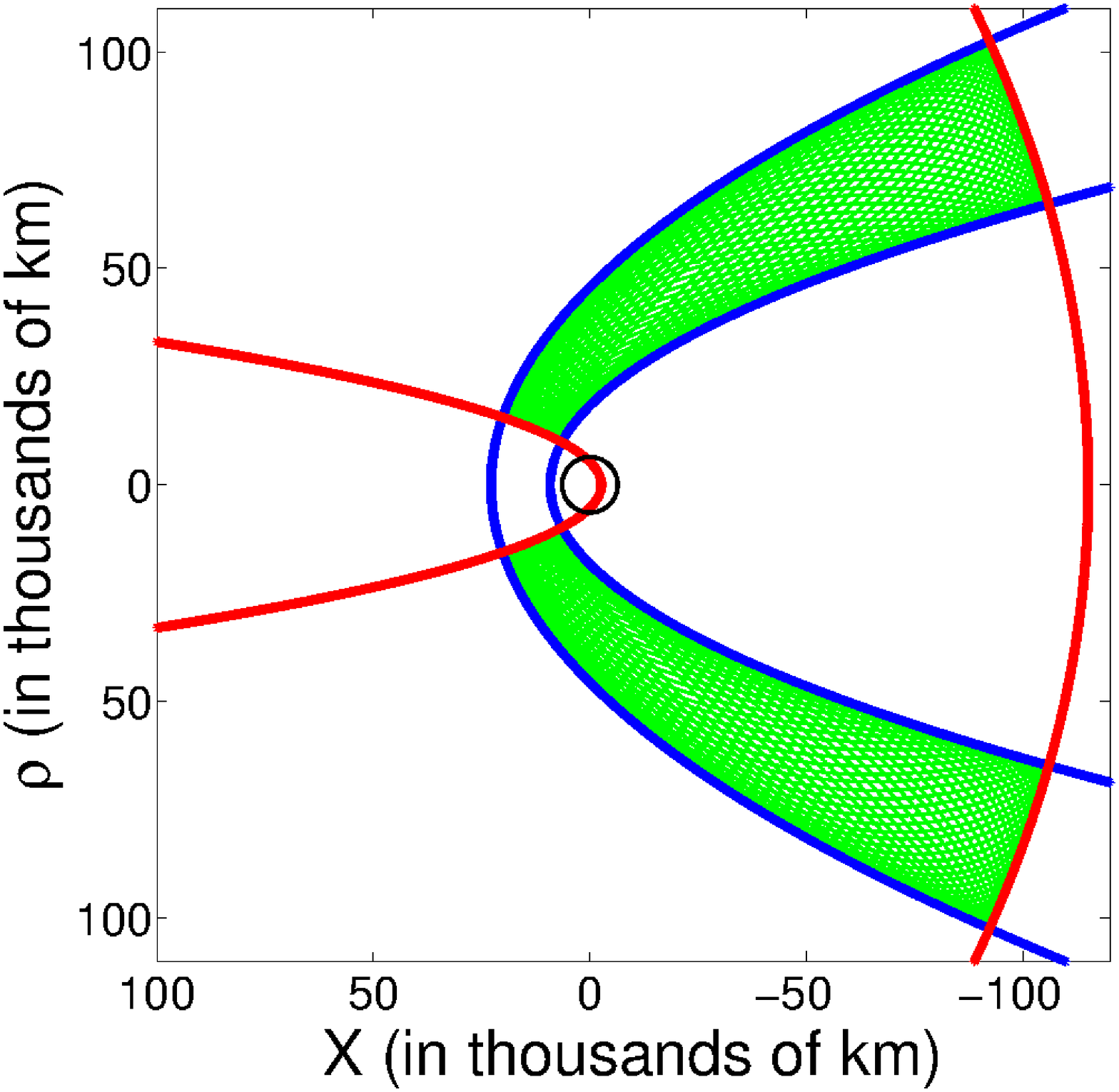}
\caption{Left panel: Combination of both panels in the figure \ref{restriction1}. The final allowed region is in green. Right panel: representation in green of the trajectory in the figure \ref{trajectoire} in the $(x,\rho)$ frame at Earth.}
\label{restriction2}
\end{figure}

As previously demonstrated, the motion is constrained in specific areas of the 3D space. The motion is always between two paraboloids due to restrictions on $u$ but it can be constrained between two other paraboloids or only one regarding $w$. Thus, the motion is constrained by four paraboloids or three, opened to infinity as shown on the figure \ref{restriction2} by the green area.

Nevertheless, there is at this point no other information on the exact motion of the particle. As shown in the figure \ref{restriction2} (right panel), the particle seems to explore all the area when its motion is restricted by four paraboloids. This is simply an observation. How can we prove that without any information on the exact motion of the particle? According to the Poincaré recurrence theorem, if the dynamical trajectory of the autonomous system evolves in a finite volume of the phase space, then in any domain, as small as it could be, there are at least two points which belong to the same trajectory. Here, the motion is constrained in space with the four paraboloids but also in velocity because $p_u$ and $p_w$ are finite values and $p_{\phi}$ is constant. All the positions in this part of the phase space can belong to the same trajectory. This is also linked to the KAM theory: here, the perturbation (the radiation pressure) affects the periodic motion (the ellipse, bounded trajectories) but it can remain quasi-periodic. As we will see, the global motion is not periodic but $u$ and $w$ motions possess their own period according to another parameter. The global motion can be periodic only if all periods are commensurable.

This is an important result because if the particle belongs to an area crossing the exobase and if this area is closed in the phase space, along a finite time, the particle will again cross the exobase. We can now define the ballistic and satellite particles as follows : both populations have no elliptic trajectories due to the radiation pressure, but they evolve in a closed domain. Depending on their constants of the motion, we can easily determine whether they cross (i.e. the domain crosses) the exobase or not, corresponding to ballistic and satellite particles respectively. Escaping particles are in the case where the initial value of $w$ is higher than the highest real root of $Q_3$ and their available area is opened to the infinity. Thus, the theorem cannot be applied here. Even if the restriction area crosses the exobase, the particle may come from infinity, come close to the exobase and go away without crossing the exobase. We need, in order to identify escaping particles (that cross the exobase and go to the infinity), to track along the time the particle to know if these particles come from the exobase or not. Thus, it is necessary to solve their trajectory along the time.

\subsection{Dimensionless expressions}\label{dimensionless_section}
For convenience, as usual in fluid mechanics, we define characteristic quantities. We decide to write all equations with dimensionless parameters. First, for distance, we define:
\begin{equation}
R_{pressure}=\sqrt{\dfrac{GM}{a}}
\end{equation}
This characteristic value was introduced by \citet{Bishop1991} and defines the limit distance where the radiation pressure overwhelms the gravitation of the planet. Then, we rewrite:
\begin{equation}
\begin{array}{lcr}
u&=&UR_{pressure}\\
w&=&WR_{pressure}
\end{array}
\end{equation}
where $U$ and $W$ are the dimensionless quantities associated to $u$ and $w$.
The energy $E$ is adimensionned with the use of the characteristic energy $k_BT_{exo}$. For $p_u$ and $p_w$, we express them in units of $\sqrt{mk_BT_{exo}}$, whereas $p_{\phi}$ is expressed in units of $\sqrt{mk_BT_{exo}}R_{pressure}$. Finally, we choose $mk_BT_{exo}\sqrt{GM/a}$ as the unit for $A$  and the unit for the adimensionless time $\tau$ is derived from the units of $u/w$ and $p_u/p_w$. We summarize these changes in table \ref{dimensionless}.

\begin{table}[!h]
\centering
\begin{tabular}{|c|c|c|}
\hline
\rule[-1ex]{0ex}{4ex}\text{dimensional}&\text{unit}&\text{dimensionless}\\
\rule[-1ex]{0ex}{2ex}\text{parameters}&&\text{parameters}\\
\hline
\hline
\rule[-1.2ex]{0ex}{4ex}$u$\  \text{and}\  $w$&$\sqrt{GM/a}=R_{pressure} $& $U$\ \text{and}\  $W$\\
\rule[-1.2ex]{0ex}{4ex}$p_u$\  \text{and}\  $p_w$&$\sqrt{mk_BT_{exo}} $& $P_U$\ \text{and}\  $P_W$\\
\rule[-1.2ex]{0ex}{4ex}$p_\phi$&$\sqrt{mk_BT_{exo}GM/a}\ $& $P_\phi$\\
\rule[-1.2ex]{0ex}{4ex}$E$&$k_BT_{exo}$&$\mathcal{E}$\\
\rule[-1.2ex]{0ex}{4ex}$A$&$mk_BT_{exo}\sqrt{GM/a}$&$\mathcal{A}$\\
\rule[-1.2ex]{0ex}{4ex}$V_u$\ \text{and}\ $V_w$&$k_BT_{exo}$&$\mathcal{V}_U$\ \text{and}\ $\mathcal{V}_W$\\
\rule[-1.2ex]{0ex}{4ex}$t$&$\sqrt{\dfrac{GMm}{ak_BT_{exo}}}$&$\mathcal{\tau}$\\
\hline
\end{tabular}
\caption{Compilation of the transformations of the parameters into dimensionless ones}

\label{dimensionless}
\end{table}

We express all the previous equations as a function of these new quantities:

\begin{equation}
\begin{array}{ccl}
\mathcal{E}&=&\dfrac{2UP^{2}_U+2WP^{2}_W}{U+W}+\dfrac{P^{2}_{\phi}}{2UW}-\dfrac{2\lambda_a}{U+W}+\dfrac{\lambda_a}{2}(U-W)
\end{array}
\label{equation357}
\end{equation}
\begin{equation}
\begin{array}{ccl}
\mathcal{A}&=&2\mathcal{E}U-4UP^{2}_{U}-\dfrac{P^{2}_{\phi}}{U}+2\lambda_a-\lambda_a U^2\\\\
&=&-2\mathcal{E}W+4WP^{2}_{W}+\dfrac{P^{2}_{\phi}}{W}-2\lambda_a-\lambda_a W^2
\end{array}
\label{equation358}
\end{equation}
\newline
\begin{equation}
\begin{array}{ccl}
P_{3}(U)&=&\lambda_a U^3-2\mathcal{E}U^2+(\mathcal{A}-2\lambda_a)U+P^{2}_{\phi}\\
&=&\lambda_a(U-U_{0})(U-U_{-})(U-U_{+})\\\\
Q_{3}(W)&=&\lambda_a W^3+2\mathcal{E}W^2+(\mathcal{A}+2\lambda_a)W-P^{2}_{\phi}\\
&=&\lambda_a(W-W_{-})(W-W_{+})(W-W_{0})
\end{array}
\label{equation359}
\end{equation}
\newline
\begin{equation}
\begin{array}{ccl}
\displaystyle{\mathcal{V}_U(U)}&=&\displaystyle{\frac{P_{\phi}^2}{2U^2}-\frac{2\lambda_a-\mathcal{A}}{2U}+\frac{\lambda_a U}{2}}\\\\
\displaystyle{\mathcal{V}_W(W)}&=&\displaystyle{\frac{P_{\phi}^2}{2W^2}-\frac{2\lambda_a+\mathcal{A}}{2W}-\frac{\lambda_a W}{2}}
\end{array}
\end{equation}
with $\lambda_a$:
\begin{equation}
\lambda_a=\dfrac{\sqrt{GMa}m}{k_{B}T}=\dfrac{GMm}{k_B T R_{pressure}}=\lambda(R_{pressure})
\end{equation}
the Jeans parameter at the distance $R_{pressure}$.

We can also reduce the equations of the motion. We introduce the dimensionless time $\tau$.

\begin{equation}
\left\{
\begin{array}{ccc}
\left(\dfrac{\mathrm{d}U}{\mathrm{d}\tau}\right)^2&=&-\dfrac{4P_{3}(U)}{(U+W)^2}\\\\
\left(\dfrac{\mathrm{d}W}{\mathrm{d}\tau}\right)^2&=&\dfrac{4Q_{3}(W)}{(U+W)^2}\\\\
\dfrac{\mathrm{d}\phi}{\mathrm{d}\tau}&=&\dfrac{P_{\phi}}{UW}
\end{array}
\right.
\end{equation}

\section{Dynamical trajectories}\label{section3}

In this part, we give implicit expressions for the dynamical trajectories of the particles, under the influence of both gravity and radiation pressure. Such expressions were already given for the 2D case with $P_{\phi}=0$ and partially generalized to the 3D case in \citet{Lantoine2011} with Jacobi elliptic functions (only bounded) \citep{Jacobi1827} and in \citet{Biscani2014} with the Weierstrass functions for bounded and unbounded trajectories. This last work dealt with the dynamical trajectories of artificial satellites but it can be applied to exospheric species subject to the radiation pressure. We propose here corrections as well as a better way to give ``simple" expressions for dynamical trajectories. We also provide expressions for unbounded trajectories that are missing in the literature. In the same way, we introduce our notations for the next papers to be published, where the influence of the radiation pressure on the density profiles and escape flux will be investigated.

According to the previous part, we have different restrictions on the motion and thus, we must distinguish the cases, that will correspond to the different types of possible trajectories. We may thus define the ballistic/satellite/escaping populations based on the roots of the $P_3$ and $Q_3$ polynoms. The trajectories are not elliptic or hyperbolic at all but one can keep their basic definitions: crossing twice the exobase for ballistic particles, orbiting but not crossing the exobase for satellite particles, coming from the exobase and escaping to the infinity for escaping particles. 

The number of real roots is a key parameter for the analytical resolution of the trajectories. The previous equations of the motion, expressed as a function of time, must be rewritten as a function of a new variable $\mathcal{T}_0$ (the subscript 0 is necessary to distinguish from $\tau$) defined as:
\begin{equation}
(U+W)\,\mathrm{d}\mathcal{T}_0=\mathrm{d}\tau
\end{equation}

Until we know the expression of $U(\mathcal{T}_0)$ and $W(\mathcal{T}_0)$, we cannot express solutions as a function of time. The system of equations leads now to:
\begin{equation}
\left\{
\begin{array}{ccc}
\left(\dfrac{\mathrm{d}U}{\mathrm{d}\mathcal{T}_0}\right)^2&=&-4P_{3}(U)\\\\
\left(\dfrac{\mathrm{d}W}{\mathrm{d}\mathcal{T}_0}\right)^2&=&4Q_{3}(W)\\\\
\dfrac{\mathrm{d}\phi}{\mathrm{d}\mathcal{T}_0}&=&P_{\phi}\left(\dfrac{1}{U}+\dfrac{1}{W}\right)\\\\
\mathrm{d}\tau&=&(U+W)\,\mathrm{d}\mathcal{T}_0\\
\end{array}
\right.
\end{equation}

All solutions, trajectories and time, are parametrized according to the variable $\mathcal{T}_0$ without physical sense.

\subsection{The U-motion}
We solve the differential equation describing the $u$-motion:
\begin{equation}
\left(\dfrac{\mathrm{d}U}{\mathrm{d}\mathcal{T}_0}\right)^2=-4P_{3}(U)=-4\lambda_a(U-U_{0})(U-U_{-})(U-U_{+})
\end{equation}
with $U_{0}$, $U_{-}$ and $U_{+}$ being real roots.

First, we set $Y^2=U-U_{0}$. The motion occurs always with $U>0$ and $U_{0}<0$ to justify this change. We obtain:
\begin{equation}
\begin{array}{rcr}
\left(\dfrac{\mathrm{d}U}{\mathrm{d}\mathcal{T}_0}\right)^2&=&-4\lambda_a(U-U_{0})(U-U_{-})(U-U_{+})\\\\
4Y^2\left(\dfrac{\mathrm{d}Y}{\mathrm{d}\mathcal{T}_0}\right)^2&=&-4\lambda_aY^2(Y^2\underbrace{-U_{-}+U_{0}}_{<0})(Y^2\underbrace{-U_{+}+U_{0}}_{<0})\\\\
\left(\dfrac{\mathrm{d}Y}{\mathrm{d}\mathcal{T}_0}\right)^2&=&-\lambda_a(Y^2-U_{-}+U_{0})(Y^2-U_{+}+U_{0})\\\\
\left(\dfrac{\mathrm{d}Y}{\sqrt{\lambda_a}\mathrm{d}\mathcal{T}_0}\right)^2&=&-(Y^2-U_{-}+U_{0})(Y^2-U_{+}+U_{0})\\\\
\end{array}
\end{equation}

Now, we set $Z=\dfrac{Y}{\sqrt{U_{+}-U_{0}}}$.

\begin{equation}
\begin{array}{rcr}
\left(\dfrac{\mathrm{d}Z}{\sqrt{\lambda_a}\mathrm{d}\mathcal{T}_0}\right)^2&=&(1-Z^2)(Z^2(U_{+}-U_{0})-U_{-}+U_{0})\\\\
\end{array}
\end{equation}

Finally, we have:
\begin{equation}
\left(\dfrac{\mathrm{d}Z}{\sqrt{\lambda_a}\sqrt{U_{+}-U_{0}}\mathrm{d}\mathcal{T}_0}\right)^2=(1-Z^2)\left(Z^2-\underbrace{\dfrac{U_{-}-U_{0}}{U_{+}-U_{0}}}_{<1\ \text{and}\ >0}\right)
\end{equation}
We define $k_U$ as:
\begin{equation}
\dfrac{U_{-}-U_{0}}{U_{+}-U_{0}}=1-\underbrace{\dfrac{U_{+}-U_{-}}{U_{+}-U_{0}}}_{<1\ \text{and}\ >0}=1-k_{U}^2
\end{equation}
\begin{equation}
k_U=\sqrt{\dfrac{U_{+}-U_{-}}{U_{+}-U_{0}}}
\end{equation}

The final equation is:

\begin{equation}
\left(\dfrac{\mathrm{d}Z}{\sqrt{\lambda_a}\sqrt{U_{+}-U_{0}}\mathrm{d}\mathcal{T}_0}\right)^2=(1-Z^2)\left(Z^2-(1-k_{U}^2)\right)
\end{equation}

The solution of this equation is:

\begin{equation}
Z=\dn(\sqrt{\lambda_a}\sqrt{U_{+}-U_{0}}(\mathcal{T}_0-\mathcal{T}_U),k_U)
\end{equation}
$\dn$ is a Jacobi elliptic function and $\mathcal{T}_U$ depends on initial conditions.

The final expression for $U$ is:
\begin{equation}
\left\{
\begin{array}{l}
U(\mathcal{T}_0)\\\\
=U_{0}+(U_{+}-U_{0})\dn^2(\sqrt{\lambda_a}\sqrt{U_{+}-U_{0}}(\mathcal{T}_0-\mathcal{T}_U),k_U)\\\\
=U_{+}-(U_{+}-U_{-})\sn^2(\sqrt{\lambda_a}\sqrt{U_{+}-U_{0}}(\mathcal{T}_0-\mathcal{T}_U),k_U)\\\\
=U_{-}+(U_{+}-U_{-})\cn^2(\sqrt{\lambda_a}\sqrt{U_{+}-U_{0}}(\mathcal{T}_0-\mathcal{T}_U),k_U)\\
\end{array}
\right.
\label{umotion}
\end{equation}
with $\cn$ and $\sn$ other Jacobi elliptic functions. These functions are $4K(k_U)$-periodic (see Appendix, eq. \ref{goodformule}).
To determine $\mathcal{T}_U$, we preferably choose the second expression for the sign.

The initial conditions give us $U(0)$, then:
\begin{equation}
\pm\arcsn\left(\sqrt{\dfrac{U_{+}-U(0)}{U_{+}-U_{-}}},k_U\right)=\sqrt{\lambda_a}\sqrt{U_{+}-U_{0}}\mathcal{T}_U
\end{equation}
where $\arcsn$ is defined as the reciprocal function of $\sn$:
\begin{equation}
\arcsn(x,k)=\int_{0}^{x}\!\dfrac{1}{\sqrt{1-t^2}\sqrt{1-k^2t^2}}\,\mathrm{d}t=F(\arcsin(x),k)
\end{equation}
with $F$ the incomplete elliptic integral of the first kind (see Appendix, eq. \ref{goodformule}).

The sign depends on $P_{U}$:
\begin{equation}
\left(\dfrac{\mathrm{d}U}{\mathrm{d}\mathcal{T}_0}\right)=4UP_U
\end{equation}
If $P_{U}$ is positive (resp. negative), $U$ increases (resp. decreases) and $-\sn^2$ increases (resp. decreases) too. Then, $\mathcal{T}_U$ is positive (resp. negative). This solution corresponds to the solution $\eta$ given by \citet{Lantoine2011}.
The $U$-motion is always constrained. The characteristics of the motion/particle (ballistic, satellite or escaping) is thus determined by conditions on $Q_3(W)$.

\subsection{The W-motion}
For the $w$-motion, we need to solve the differential equation:
\begin{equation}
\left(\dfrac{\mathrm{d}W}{\mathrm{d}\mathcal{T}_0}\right)^2=4Q_{3}(W)=4\lambda_a(W-W_{0})(W-W_{-})(W-W_{+})
\end{equation}
$W_0$ is a positive real root but we must distinguish between various cases for the other roots: $W_{+}$ and $W_{-}$ may be real or complex conjugate roots.

\subsubsection{Three real roots}
We must consider two cases: $W>W_{0}$ and $W_{+}>W>W_{-}$. The second one occurs mathematically and not only physically when $W_{-}$ and $W_{+}$ are positive. 

\paragraph{a) For $W_{-}<W<W_{+}$}
 This case occurs for bounded trajectories (ballistic or satellite trajectory). Here, this is the same treatment as for the $U$-motion. After setting the two following substitutions $Y^2=W_0-W$ and $Z=\dfrac{Y}{\sqrt{W_{0}-W_{-}}}$, we obtain a similar solution:





\begin{equation}
Z=\dn(\sqrt{\lambda_a}\sqrt{W_{0}-W_{-}}(\mathcal{T}_0-\mathcal{T}_W),k_W)
\end{equation}

The final expression for $W$ is
\begin{equation}
\left\{
\begin{array}{l}
W(\mathcal{T}_0)\\\\
=W_{0}-(W_{0}-W_{-})\dn^2(\sqrt{\lambda_a}\sqrt{W_{0}-W_{-}}(\mathcal{T}_0-\mathcal{T}_W),k_W)\\\\
=W_{+}-(W_{+}-W_{-})\cn^2(\sqrt{\lambda_a}\sqrt{W_{0}-W_{-}}(\mathcal{T}_0-\mathcal{T}_W),k_W)\\\\
=W_{-}+(W_{+}-W_{-})\sn^2(\sqrt{\lambda_a}\sqrt{W_{0}-W_{-}}(\mathcal{T}_0-\mathcal{T}_W),k_W)\\
\end{array}
\right.
\label{wmotion1}
\end{equation}
The initial conditions giving us $W(0)$ and then leads to:
\begin{equation}
\pm\arcsn\left(\sqrt{\dfrac{W(0)-W_-}{W_{+}-W_{-}}},k_W\right)=\sqrt{\lambda_a}\sqrt{W_{0}-W_{-}}\mathcal{T}_U
\end{equation}

The sign depends on $P_{W}$: 
\begin{equation}
\left(\dfrac{\mathrm{d}W}{\mathrm{d}\mathcal{T}_0}\right)=4WP_W
\end{equation}
If $P_{W}$ is positive (resp. negative), $W$ increases (resp. decreases) and $\sn^2$ increases (resp. decreases) too. Then, $\mathcal{T}_W$ is negative (resp. positive). This solution corresponds to the solution $\xi I$ given by \citet{Lantoine2011}.

\paragraph{b) For $W>W_{0}$}

First, the motion occurs only for $W\in[W_0;+\infty[$ (corresponding to escaping particles). Then, we set $Y^2=W-W_{0}$. We obtain:

\begin{equation}
\begin{array}{rcr}
\left(\dfrac{\mathrm{d}W}{\mathrm{d}\mathcal{T}_0}\right)^2&=&4\lambda_a(W-W_{0})(W-W_{-})(W-W_{+})\\\\
4Y^2\left(\dfrac{\mathrm{d}Y}{\mathrm{d}\mathcal{T}_0}\right)^2&=&4\lambda_aY^2(Y^2\underbrace{-W_{-}+W_{0}}_{>0})(Y^2\underbrace{-W_{+}+W_{0}}_{>0})\\\\
\left(\dfrac{\mathrm{d}Y}{\mathrm{d}\mathcal{T}_0}\right)^2&=&\lambda_a(Y^2-W_{-}+W_{0})(Y^2-W_{+}+W_{0})\\\\
\left(\dfrac{\mathrm{d}Y}{\sqrt{\lambda_a}\mathrm{d}\mathcal{T}_0}\right)^2&=&(Y^2-W_{-}+W_{0})(Y^2-W_{+}+W_{0})\\\\
\end{array}
\end{equation}

Now, we set $Z=\dfrac{Y}{\sqrt{W_{0}-W_{-}}}$.

Finally, we have:
\begin{equation}
\left(\dfrac{\mathrm{d}Z}{\sqrt{\lambda_a}\sqrt{W_{0}-W_{-}}\mathrm{d}\mathcal{T}_0}\right)^2=(Z^2+1)\left(Z^2+\underbrace{\dfrac{W_{0}-W_{+}}{W_{0}-W_{-}}}_{<1\ \text{and}\ >0}\right)
\end{equation}
We define $k_W$ as:
\begin{equation}
\dfrac{W_{0}-W_{+}}{W_{0}-W_{-}}=1-\underbrace{\dfrac{W_{+}-W_{-}}{W_{0}-W_{-}}}_{<1\ \text{and}\ >0}=1-k_{W}^2
\end{equation}
\begin{equation}
k_W=\sqrt{\dfrac{W_{+}-W_{-}}{W_{0}-W_{-}}}
\end{equation}

The solution of this equation is
\begin{equation}
Z=\cs(\sqrt{\lambda_a}\sqrt{W_{0}-W_{-}}(\mathcal{T}_0-\mathcal{T}_W),k_W)
\end{equation}
$\cs$ is a Jacobi elliptic function and $\mathcal{T}_W$ depends on the initial conditions.

The final expression for $W$ is:
\begin{equation}
\left\{
\begin{array}{l}
W(\mathcal{T}_0)\\\\
=W_{0}+(W_{0}-W_{-})\cs^2(\sqrt{\lambda_a}\sqrt{W_{0}-W_{-}}(\mathcal{T}_0-\mathcal{T}_W),k_W)\\\\
=W_{+}+(W_{0}-W_{-})\ds^2(\sqrt{\lambda_a}\sqrt{W_{0}-W_{-}}(\mathcal{T}_0-\mathcal{T}_W),k_W)\\\\
=W_{-}+(W_{0}-W_{-})\ns^2(\sqrt{\lambda_a}\sqrt{W_{0}-W_{-}}(\mathcal{T}_0-\mathcal{T}_W),k_W)\\
\end{array}
\right.
\label{wmotion2}
\end{equation}

These functions are $4K(k_W)$-periodic and are defined on $\mathcal{T}_0-\mathcal{T}_W\in\mathbb{R}/\{ 4mK(k_W)|m\in\mathbb{Z}\}$. These functions diverge at $4mK(k_W)$ but this is not an issue: the motion can diverge with respect to $\mathcal{T}_0$, but, according to the time $\tau$ and the integration (see Section \ref{Time}), the $w$-motion remains continuous for $\tau\in\mathbb{R}$ as

\begin{equation*}
\mathcal{T}_0-\mathcal{T}_W\in]0;4K(k_W)[\Longleftrightarrow \mathcal{\tau}\in\mathbb{R}
\end{equation*}

Let us assume the initial conditions provide $W(0)$, then:
\begin{equation}
\pm\arccs\left(\sqrt{\dfrac{W(0)-W_{0}}{W_{0}-W_{-}}},k_W\right)=\sqrt{\lambda_a}\sqrt{W_{0}-W_{-}}\mathcal{T}_W
\end{equation}
with
\begin{equation}
\arccs(x,k)=\int_{0}^{x}\!\dfrac{1}{\sqrt{1+t^2}\sqrt{t^2-1+k^2}}\,\mathrm{d}t=F\left(\arctan\left(\dfrac{1}{x}\right),k\right)
\end{equation}

The sign depends on $P_{W}$:
\begin{equation}
\left(\dfrac{\mathrm{d}W}{\mathrm{d}\mathcal{T}_0}\right)=4WP_W
\end{equation}

If $P_{W}$ is positive (resp. negative), $W$ increases (resp. decreases) and $\cs^2$ increases (resp. decreases) too. Then, $\mathcal{T}_W$ is positive (resp. negative). This solution corresponds to the solution $\xi II$ given by \citet{Lantoine2011}.
These expressions are useful only for three real roots. A last case remains: only one real positive root.

\subsubsection{Only one real positive root}

For any initial conditions, the particle will be escaping if $Q_3$ has only one real root. The solution for this case is more complex compared with the previous solutions. As done by \citet{Lantoine2011}, we could apply some transformations and obtain a new expression, that would be a combination of $\cn$ and $\sn$. Nevertheless, we propose a simpler way to determine the expression with the knowledge of all roots, even imaginaries. We start from the equation \ref{Wdiff} with $W_{+}$ and $W_{-}$ not real:
\begin{equation}
\left(\dfrac{\mathrm{d}W}{\mathrm{d}\mathcal{T}_0}\right)^2=4\lambda_a(W-W_0)(W-W_+)(W-W_-)
\label{Wdiff}
\end{equation}

After the separation of variables, we obtain:
\begin{equation}
\int^W\!\dfrac{\mathrm{d}W'}{\sqrt{(W'-W_0)(W'-W_+)(W'-W_-)}}=2\sqrt{\lambda_a}(\mathcal{T}_0-\mathcal{T}_W)
\label{equation60}
\end{equation}

Now, we apply the procedure proposed in \cite{abra} (p. 597) in the case where we have only one real root. First, we define:
\begin{equation}
\lambda^{2}=\sqrt{(W_0-W_+)(W_0-W_-)}=\sqrt{(W_0-\operatorname{Re}(W_+))^2+\operatorname{Im}(W_+)^2}=\sqrt{\dfrac{Q_{3}^{'}(W_0)}{\lambda_a}}
\end{equation}

and also
\begin{equation}
k_{W_0}=\sqrt{\dfrac{1}{2}-\dfrac{1}{4}\dfrac{W_0-W_++W_0-W_-}{\lambda^2}}=\sqrt{\dfrac{1}{2}-\dfrac{1}{2}\dfrac{W_0-\operatorname{Re}(W_+)}{\lambda^2}}
\end{equation}

According to \cite{abra} (p. 597), the left hand side corresponds to:
\begin{equation}
\int^W_{W_0}\!\dfrac{\mathrm{d}W'}{\sqrt{(W'-W_0)(W'-W_+)(W'-W_-)}}=\dfrac{F(\theta,k_{W_0})}{\lambda}
\end{equation}

with $F$ the Elliptic function of first kind (defined in the appendix \ref{appendixA}) and
\begin{equation}
\cos\theta=\dfrac{\lambda^2-(W-W_0)}{\lambda^2+(W-W_0)} \Longrightarrow W=W_0+\lambda^2\dfrac{1-\cos\theta}{1+\cos\theta}
\end{equation}

According to the equation \ref{equation60}:
\begin{equation}
\theta=\am(2\lambda\sqrt{\lambda_a}(\mathcal{T}_0-\mathcal{T}_{W_0}),k_{W_0})
\end{equation}

with $\am$ the Jacobi amplitude defined as
\begin{equation}
\am(x,k)=F^{-1}(x,k)
\end{equation}

Thus
\begin{equation}
W(\mathcal{T}_0)=W_0+\lambda^2\dfrac{1-\cn(2\lambda\sqrt{\lambda_a}(\mathcal{T}_0-\mathcal{T}_{W_0}),k_{W_0})}{1+\cn(2\lambda\sqrt{\lambda_a}(\mathcal{T}_0-\mathcal{T}_{W_0}),k_{W_0})}
\end{equation}

Finally, one can simplify the expression based on \cite{abra2} (equation 22.6.18):

\begin{equation}
W(\mathcal{T}_0)=W_0+\lambda^2\dfrac{\dn^2(\lambda\sqrt{\lambda_a}(\mathcal{T}_0-\mathcal{T}_{W_0}),k_{W_0})}{\cs^2(\lambda\sqrt{\lambda_a}(\mathcal{T}_0-\mathcal{T}_{W_0}),k_{W_0})}
\label{wmotion3}
\end{equation}

This function is $4K(k_{W_0})$-periodic and is defined on $\mathcal{T}_0-\mathcal{T}_W\in\mathbb{R}/\{ 2K(k_{W_0})+4mK(k_{W_0})|m\in\mathbb{Z}\}$. This function diverges at $2K(k_{W_0})+4mK(k_{W_0})$ but this does not impact our result: the motion can diverge with respect to $\mathcal{T}_0$, but, according to the time $\tau$ and the integration (see Section \ref{Time}), the $w$-motion remains continuous for $\tau\in\mathbb{R}$ as

\begin{equation*}
\mathcal{T}_0-\mathcal{T}_{W_0}\in]-2K(k_{W_0});2K(k_{W_0})[\Longleftrightarrow \mathcal{\tau}\in\mathbb{R}
\end{equation*}

As usual, according to the initial conditions, we define $\mathcal{T}_{W_0}$ as
\begin{equation}
\pm\arccn\left(\dfrac{\lambda^2-(W(0)-W_0)}{\lambda^2+(W(0)-W_0)},k_{W_0}\right)=2\lambda\sqrt{\lambda_a}\mathcal{T}_{W_0}
\end{equation}
with
\begin{equation}
\arccn(x,k)=\int_{x}^{1}\!\dfrac{1}{\sqrt{1+t^2}\sqrt{1-k^2+k^2t^2}}\,\mathrm{d}t=F(\arccos(x),k)
\end{equation}

If $P_W$ is positive (resp. negative) then $\mathcal{T}_{W_0}$ is negative (resp. positive). This solution corresponds to the solution $\eta$ given by \citet{Lantoine2011} but our derived solution is less complex to use (with less time computing).

\section{Time equation}\label{Time}
We gave analytical formulations for the different kinds of trajectories, expressed implicitly. Now, since we have all expressions of the trajectories as a function of $\mathcal{T}_0$, we can express the real time $\tau$ (or $t$):
\begin{equation}
\begin{array}{ccl}
t&=&\sqrt{\dfrac{GMm}{ak_BT_{exo}}}\tau\\\\
\tau(\mathcal{T}_0)&=&\displaystyle\int_{0}^{\mathcal{T}_0}\!U(\mathcal{T}^\prime)+W(\mathcal{T}^\prime)\,\mathrm{d}\mathcal{T}^\prime
\end{array}
\end{equation}

Here, we use MATHEMATICA and MAPLE to derive the primitives. It is necessary to be very careful since these different programs can have different definitions of the elliptic functions for example. To avoid these issues, we remind at each use the definition employed. The first part of the integral gives:
\begin{equation}
\begin{array}{l}
\displaystyle\int_{0}^{\mathcal{T}_0}\!U(\mathcal{T}^\prime)\,\mathrm{d}\mathcal{T}^\prime\\\\
=U_{0}\mathcal{T}_0\\\\
+\dfrac{\alpha}{\lambda_a}[E(\am(\alpha(\mathcal{T}_0-\mathcal{T}_U),k_U),k_U)-E(\am(-\alpha\mathcal{T}_U,k_U),k_U)]
\end{array}
\label{tu}
\end{equation}
with $\alpha=\sqrt{\lambda_a}\sqrt{U_+-U_0}$, $E$ the incomplete elliptic function of the second kind (see Appendix, eq. \ref{goodformule}).

The second part of the integral is more complex because we have different expressions according to the number of roots and the initial conditions. In the case of three real roots, if $W_+>W(0)>W_-$ then
\begin{equation}
\begin{array}{l}
\displaystyle\int_{0}^{\mathcal{T}_0}\!W(\mathcal{T}^\prime)\,\mathrm{d}\mathcal{T}^\prime\\\\
=W_{0}\mathcal{T}_0\\\\
-\dfrac{\beta}{\lambda_a}[E(\am(\beta(\mathcal{T}_0-\mathcal{T}_W),k_W),k_W)-E(\am(-\beta\mathcal{T}_W,k_W),k_W]
\end{array}
\label{tw1}
\end{equation}
with $\beta=\sqrt{\lambda_a}\sqrt{W_0-W_-}$.\newline

If $W(0)>W_0$ with three real roots then
\begin{equation}
\begin{array}{l}
\displaystyle\int_{0}^{\mathcal{T}_0}\!W(\mathcal{T}^\prime)\,\mathrm{d}\mathcal{T}^\prime\\\\
=W_{0}\mathcal{T}_0\\\\
-\dfrac{\beta}{\lambda_a}[E(\am(\beta(\mathcal{T}_0-\mathcal{T}_W),k_W),k_W)-E(\am(-\beta\mathcal{T}_W,k_W),k_W)]\\\\
-\dfrac{\beta}{\lambda_a}\left[\dfrac{\cn(\beta(\mathcal{T}_0-\mathcal{T}_W),k_W)}{\sd(\beta(\mathcal{T}_0-\mathcal{T}_W),k_W)}
-\dfrac{\cn(-\beta\mathcal{T}_W,k_W)}{\sd(-\beta\mathcal{T}_W,k_W)}\right]
\end{array}
\label{tw2}
\end{equation}

Finally, in the case of only one real root, the time equation is given by:
\begin{equation}
\begin{array}{l}
\displaystyle\int_{0}^{\mathcal{T}_0}\!W(\mathcal{T}^\prime)\,\mathrm{d}\mathcal{T}^\prime\\\\
=(W_{0}+\lambda^2)\mathcal{T}_0\\\\
-\dfrac{\gamma}{\lambda_a}[E(\am(2\gamma(\mathcal{T}_0-\mathcal{T}_{W_0}),k_{W_0}),k_{W_0})-E(\am(-2\gamma\mathcal{T}_{W_0},k_{W_0}),k_{W_0})]\\\\
+\dfrac{\gamma}{\lambda_a}\left[\dfrac{\sn(2\gamma(\mathcal{T}_0-\mathcal{T}_{W_0}),k_{W_0})\dn(2\gamma(\mathcal{T}_0-\mathcal{T}_{W_0}),k_{W_0})}{1+\cn(2\gamma(\mathcal{T}_0-\mathcal{T}_{W_0}),k_{W_0})}
-\dfrac{\sn(-2\gamma\mathcal{T}_{W_0},k_{W_0})\dn(-2\gamma\mathcal{T}_{W_0},k_{W_0})}{1+\cn(-2\gamma\mathcal{T}_{W_0},k_{W_0})}\right]
\end{array}
\label{tw3}
\end{equation}
with $\gamma=\lambda\sqrt{\lambda_a}$.

For both last cases, $\mathcal{T}_0-\mathcal{T}_W$ belongs to $]0;4K(k_W)[$ (respectively $]-2K(k_{W_0});2K(k_{W_0})[$) for the equation \ref{tw2} (resp. for the equation \ref{tw3}). Nevertheless, when $\mathcal{T}_0-\mathcal{T}_W$ tends to $4K(k_{W_0})$ (resp. $2K(k_{W_0})$), the integration diverges and $\tau$ too. The $w$-motion occurs on an subset of $\mathbb{R}$ with respect to $\mathcal{T}_0$ but on $\mathbb{R}$ entirely with respect to $\tau$. The transformation of $\tau$ into $\mathcal{T}_0$ is bijective because the integrand $U+W$ is strictly positive.
\section{$\phi$-equation}

To complete the description of the motion as a function of time, it is also necessary to solve the evolution of the angle $\phi$, obeying to:
\begin{equation}
\begin{array}{ccl}
\dfrac{\mathrm{d}\phi}{\mathrm{d}\mathcal{T}_0}&=&P_{\phi}\left(\dfrac{1}{U(\mathcal{T}_0)}+\dfrac{1}{W(\mathcal{T}_0)}\right)\\\\
\phi(\mathcal{T}_0)-\phi(0)&=&\displaystyle\int_{0}^{\mathcal{T}_0}\! P_{\phi}\left(\dfrac{1}{U(\mathcal{T}^\prime)}+\dfrac{1}{W(\mathcal{T}^\prime)}\right)\,\mathrm{d}\mathcal{T}^\prime
\end{array}
\end{equation}
 As already done in the previous part, we separate into two integrals. The first part still gives:
\begin{equation}
\begin{array}{l}
\displaystyle\int_{0}^{\mathcal{T}_0}\!\dfrac{1}{U(\mathcal{T}^\prime)}\,\mathrm{d}\mathcal{T}^\prime\\\\
=\dfrac{\Pi\left(1-\dfrac{U_{-}}{U_{+}};\am(\alpha(\mathcal{T}_0-\mathcal{T}_U),k_U),k_U\right)}{\alpha U_{+}}\\\\
-\dfrac{\Pi\left(1-\dfrac{U_{-}}{U_{+}};\am(-\alpha\mathcal{T}_U,k_U),k_U\right)}{\alpha U_{+}}
\end{array}
\label{phiu}
\end{equation}
with $\Pi$ the incomplete elliptic integral of the third kind (see Appendix, eq. \ref{goodformule}).\newline

In the three real roots case for $Q_3$, if initially we have $W_+>W(0)>W_-$:
\begin{equation}
\begin{array}{l}
\displaystyle\int_{0}^{\mathcal{T}'}\!\dfrac{1}{W(\mathcal{T}_0)}\,\mathrm{d}\mathcal{T}_0\\\\
=\dfrac{\Pi\left(1-\dfrac{W_{+}}{W_{-}};\am(\beta(\mathcal{T}'-\mathcal{T}_W),k_W),k_W\right)}{\beta W_{-}}\\\\
-\dfrac{\Pi\left(1-\dfrac{W_{+}}{W_{-}};\am(-\beta\mathcal{T}_W,k_W),k_W\right)}{\beta W_{-}}
\end{array}
\label{phiw1}
\end{equation}
or if initially $W(0)>W_0$:
\begin{equation}
\begin{array}{l}
\displaystyle\int_{0}^{\mathcal{T}_0}\!\dfrac{1}{W(\mathcal{T}^\prime)}\,\mathrm{d}\mathcal{T}^\prime\\\\
=\dfrac{\beta\mathcal{T}_0-\Pi\left(-\dfrac{W_{-}}{W_{0}-W_{-}};\am(\beta(\mathcal{T}_0-\mathcal{T}_W),k_W),k_W\right)}{\beta W_{-}}\\\\
+\dfrac{\Pi\left(-\dfrac{W_{-}}{W_{0}-W_{-}};\am(-\beta\mathcal{T}_W,k_W),k_W\right)}{\beta W_{-}}
\end{array}
\label{phiw2}
\end{equation}

and for the last case with one real root:
\begin{equation}
\begin{array}{l}
\displaystyle\int_{0}^{\mathcal{T}_0}\!\dfrac{1}{W(\mathcal{T}^\prime)}\,\mathrm{d}\mathcal{T}^\prime\\\\
=\dfrac{\mathcal{T}_0}{W_0-\lambda^2}\\\\
-\dfrac{(W_0+\lambda^2)}{4W_0\gamma(W_0-\lambda^2)}\Pi\left(-\dfrac{(W_0-\lambda^2)^2}{4\lambda^2W_0};\am(2\gamma(\mathcal{T}_0-\mathcal{T}_{W_0}),k_{W_0}),k_{W_0}\right)\\\\
+\dfrac{1}{2P_{\phi}}\arctan\left(\dfrac{P_{\phi}}{2\gamma W_0}\dfrac{\sn(2\gamma(\mathcal{T}_0-\mathcal{T}_{W_0}),k_{W_0})}{\dn(2\gamma(\mathcal{T}_0-\mathcal{T}_{W_0}),k_{W_0})}\right)\\\\
+\dfrac{(W_0+\lambda^2)}{4W_0\gamma(W_0-\lambda^2)}\Pi\left(-\dfrac{(W_0-\lambda^2)^2}{4\lambda^2W_0};\am(-2\gamma\mathcal{T}_{W_0},k_{W_0}),k_{W_0}\right)\\\\
-\dfrac{1}{2P_{\phi}}\arctan\left(\dfrac{P_{\phi}}{2\gamma W_0}\dfrac{\sn(-2\gamma\mathcal{T}_{W_0},k_{W_0})}{\dn(-2\gamma\mathcal{T}_{W_0},k_{W_0})}\right)\\\\
\end{array}
\label{phiw3}
\end{equation}

This last formula is only available for $\mathcal{T}_0-\mathcal{T}_{W_0} \in ]-2K(k_{W_0});2K(k_{W_0})[$, range where it is continuous.

\section{Circular orbits}\label{circular}

\begin{figure}[!h]
\centering
\includegraphics[height=0.3\linewidth]{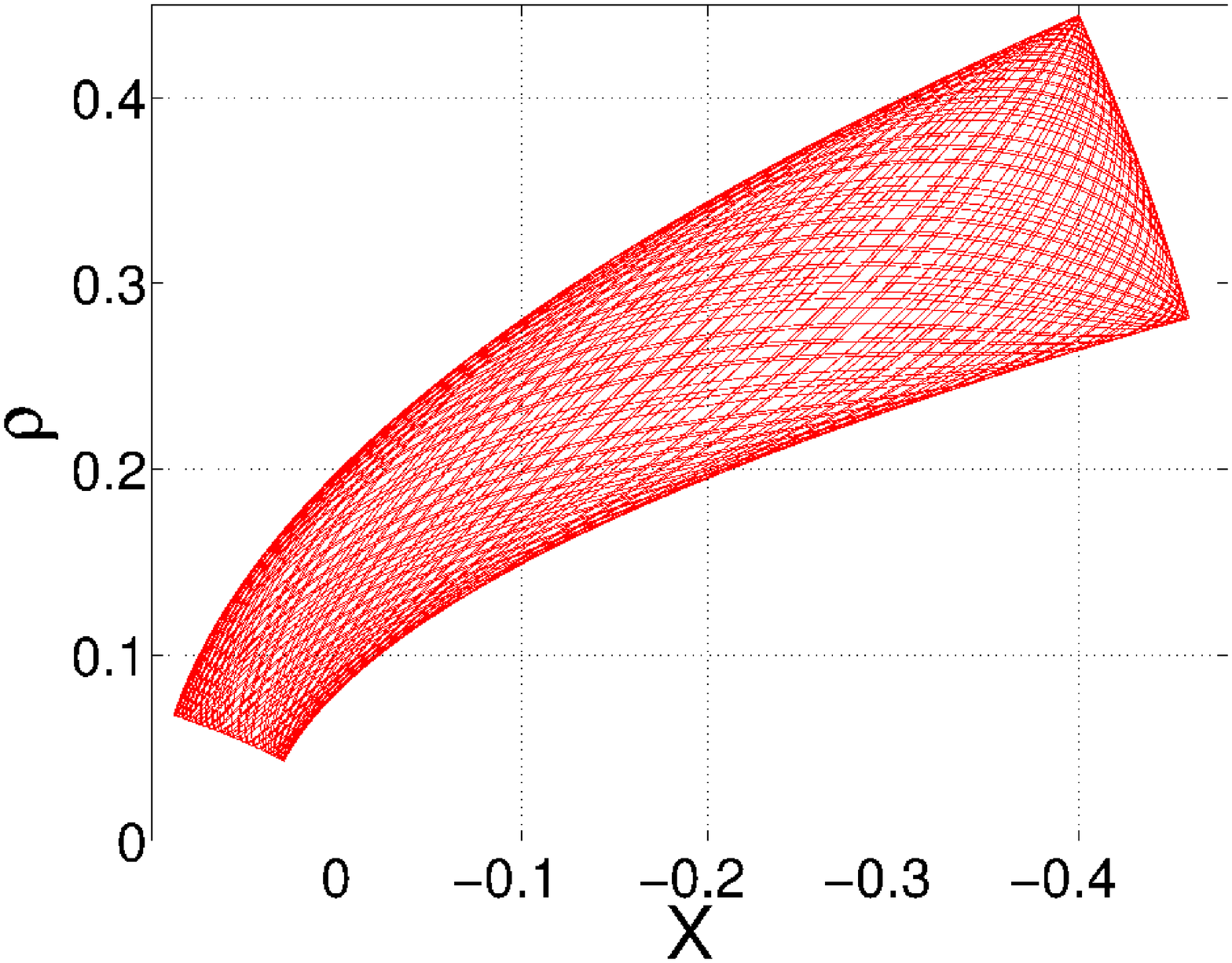}\includegraphics[height=0.30\linewidth]{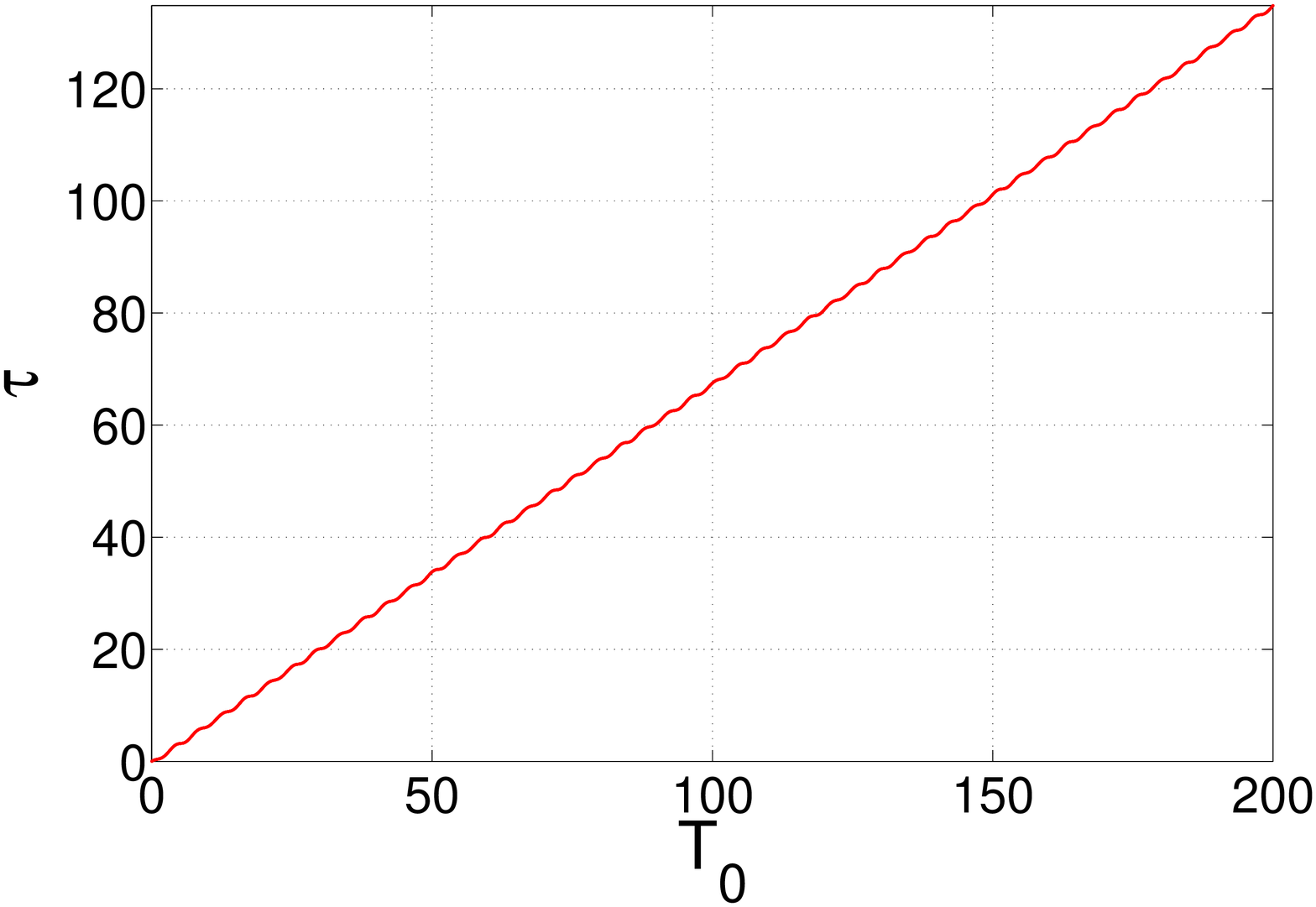}\\
\includegraphics[height=0.3\linewidth]{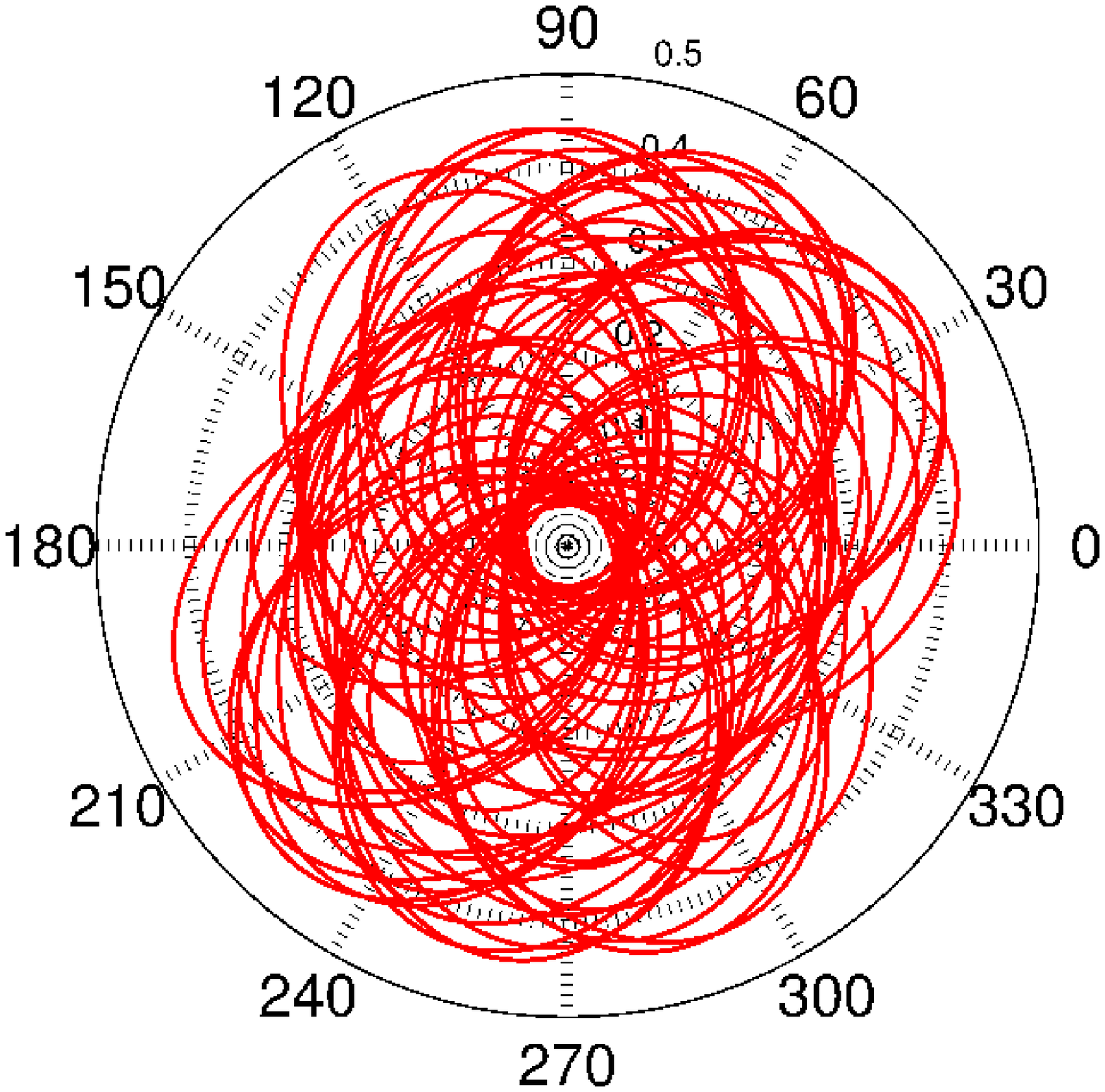}\hfill\includegraphics[height=0.3\linewidth]{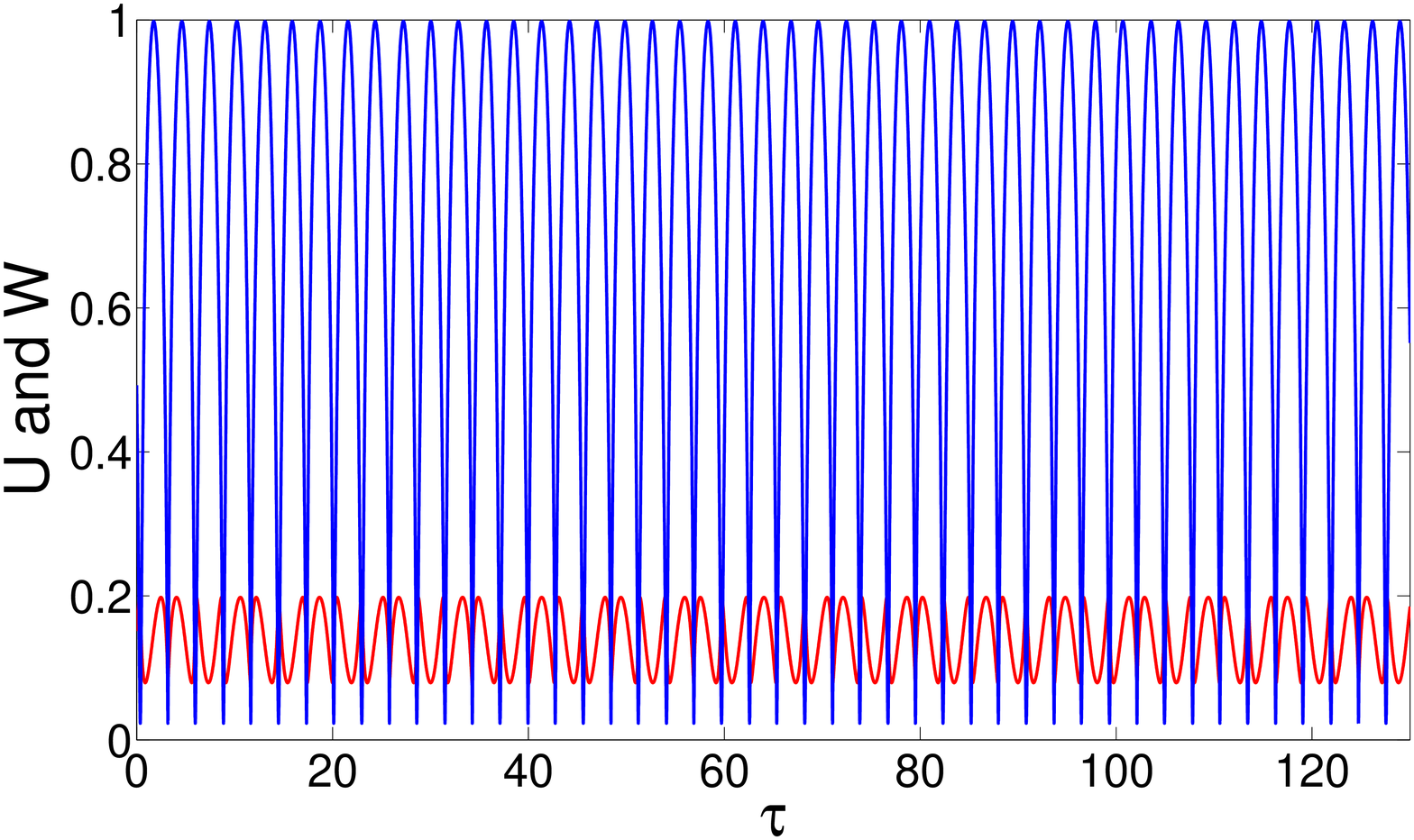}
\caption{Plots of a bounded particle (or spacecraft) motion in the $(X,\rho)$ plane (upper left panel), of the time as a function of $\mathcal{T}_0$ (upper right  panel), of the motion in polar coordinates $(\phi,\rho)$ (lower left panel) and the $U-W$ coordinates as a function of the time (red for $U$, blue for $W$, lower right panel). $U$ and $W$ do not show any periodicity because their periods are not commensurable (i.e. the ratio is a rational number) with $\mathcal{T}_0$ and thus with the time $\tau$}
\label{sample}
\end{figure}

With the solutions previously derived, we can know the exact motion of a bounded or unbounded particle as a function of the time such as given in figure \ref{sample}. It is clear that even a bounded trajectory the motion has no periodicity at all (especially for the $\phi$ motion). Nevertheless, it could be interesting to focus on stable bounded orbits and search for periodic motions (as in \citet{Biscani2014}), and thus investigate in particular the circular stable orbits for spacecrafts \citep{Namouni2007} or the possible positions for satellite particles produced by collisions in the exosphere \citep{Beth2014}. Thus, we dedicate this section to the conditions to obtain such orbits.

For a specific set of initial conditions, it can be possible to obtain circular orbits. This orbit occurs when, on the one hand, the attraction of the planet projected along the $x$-axis is equal to acceleration due to the radiation pressure:
\begin{equation}
-\dfrac{GMmx}{r^3}-ma=0
\end{equation}

 In dimensionless quantity, this will be expressed by:
\begin{equation}
R^2+\cos\theta=0
\label{R}
\end{equation}

On the other hand, it is also necessary that the centrifugal force induced by the rotation around the $x$-axis is equal to the acceleration around the planet in the perpendicular plane to the $x$-axis. Thus, we obtain the secondary equality:
\begin{equation}
-\dfrac{GMm\rho}{r^3}+\dfrac{mv_{\phi}^{2}}{\rho^2}=0
\end{equation}

 In dimensionless quantity, this will be expressed by:
\begin{equation}
\lambda_a R\sin^4\theta-P_{\phi}^{2}=0
\label{centrifuge}
\end{equation}

Combining these two equations, we obtain:
\begin{equation}
(\sin^2\theta)^9-(\sin^2\theta)^8+\dfrac{P_{\phi}^{2}}{\lambda_a}=0
\end{equation}

We need to study the polynom
\begin{equation}
P(X)=X^9-X^8+\dfrac{P_{\phi}^{2}}{\lambda_a}
\label{polynom}
\end{equation}
with $X=\sin^2\theta\in[0;1]$. Depending on the $P_{\phi}^{2}$ values, we have zero, one or two solutions for $P(X)=0$. Indeed, $P(0)=P(1)>0$ so that according to the Rolle theorem, there is $a\in]0;1[$ with $P'(a)=0$. Here, $a$ is equal to 8/9. Nevertheless, if $P(a)$ is positive then $P(X)=0$ does not have solutions and inversely, if $P(a)$ is negative then we have two solutions. This value is:
\begin{equation}
P\left(\dfrac{8}{9}\right)=\dfrac{P_{\phi}^{8}}{\lambda^{4}_{a}}-\dfrac{1}{9}\left(\dfrac{8}{9}\right)^8
\end{equation}

A critical maximum value of $P_{\phi}$ thus exists to allow for circular orbits and is 
\begin{equation}
|P_{c\phi}|=\dfrac{8\sqrt{\lambda_a}}{9\sqrt[4]{3}}
\end{equation}

Above this value, we cannot find any bounded trajectories: there is no any equilibrium point and thus no circular orbits (stable or not). For lower values of $|P_{\phi}|$, we have two solutions for $P(X)=0$: one stable and one unstable as shown by \citet{Namouni2007} and theirs plots of the equipotentials. These two solutions correspond respectively to the stable point around which the equipotentials are closed and to the saddle point, which is the last limit where one can find closed equipotentials, and is the only point where two equipotentials can cross. As long as $|P_{\phi}|<|P_{c\phi}|$, these both specific points exist: they have the same $U$. Physically, the potential $\mathcal{V}_W$ has two extrema as plotted in figure \ref{potentiel}. When $|P_{\phi}|$ reaches $|P_{c\phi}|$, the local minimum goes to the right and the local maximum goes to the left at the same location $W_{crit}$. For higher $|P_{\phi}|$ values, $V_W$ have no extremum anymore: the potential is strictly decreasing and the particles are unbounded (escaping). Thus, the bounded particles, satellite and ballistic particles, have $|P_{\phi}|<|P_{c\phi}|$. The distinction between them thus depend on if they cross or not the exobase.

 The critical values are given in $(z,\rho)$ coordinates ($z$ is $-x$ for us, in comparison with \citet{Namouni2007}). In dimensionless unities and in $(R,\theta)$ using the equations \ref{centrifuge} then \ref{R}, the critical orbit is:
\begin{equation}
(R_{crit},\theta_{crit})=\left(\dfrac{1}{\sqrt{3}},\pi-\arcsin\left(\dfrac{2\sqrt{2}}{3}\right)\right)
\end{equation}
or in $(U,W)$ coordinates:
\begin{equation}
(U_{crit},W_{crit})=\left(\dfrac{2}{3\sqrt{3}},\dfrac{4}{3\sqrt{3}}\right)
\end{equation}

The real positive roots of the polynomial \ref{polynom} combined with the equality \ref{R} give the positions of the circular orbits (two coordinates are necessary) allowed to spacecraft or particles under the influence of both gravity and radiation pressure.

\section{Summary}\label{summary}

The knowledge of the exact trajectories of particles or satellites under the influence of gravity and radiation pressure needs the calculation of the spatial coordinates, i.e. the $U/W/\phi$ motions, as well as the time evolution. We summarize all needed equations in  table \ref{summary}.

\begin{figure*}[!t]
\centering
\begin{tabular}{|c|c||c|c|c|}
\hline
\rule[-1ex]{0ex}{4ex}\text{cases}&\text{Three real roots}    &\multicolumn{2}{c|}{\text{Three real roots}}&\text{One real root}\\
\rule[-1ex]{0ex}{2ex}        &\text{for $P_3$ (always)}&\multicolumn{2}{c|}{\text{for $Q_3$}}         & \text{for $Q_3$}\\
\hline
\rule[-1ex]{0ex}{4ex} \text{trajectory}       &\text{bounded}&\text{bounded} &\text{unbounded}& \text{unbounded}\\
\hline
\hline
\rule[-1.2ex]{0ex}{4ex}\text{$U$-motion}& \ref{umotion}(L2D)&&&\\
\rule[-1.2ex]{0ex}{4ex}\text{$W$-motion}&&\ref{wmotion1}(L2D)&\ref{wmotion2}(L2D)&\ref{wmotion3}(s)\\
\rule[-1.2ex]{0ex}{4ex}\text{time equation}&\ref{tu}(L2D)&\ref{tw1}(L2D) & \ref{tw2}*NEW*&\ref{tw3}*NEW*\\
\rule[-1.2ex]{0ex}{4ex}\text{$\phi$-motion}&\ref{phiu}(L3D)&\ref{phiw1}(L3D)& \ref{phiw2}*NEW*&\ref{phiw3}*NEW*\\
\rule[-1.2ex]{0ex}{4ex}\text{$\mathcal{T}_0$ range}&$\mathcal{T}_0 \in \mathbb{R}$&$\mathcal{T}_0 \in\mathbb{R}$&$\mathcal{T}_0\in[0;(1-\text{sg}(\mathcal{T}_W))2K(k_W)+\mathcal{T}_W[$&$\mathcal{T}_0\in[0;2K(k_{W_0})+\mathcal{T}_W[$\\
\hline
\end{tabular}
\caption{Summary of the different solutions for each kind of motion. The notation *NEW* corresponds to the new solutions derived in this paper, not yet derived in previous works. The symbol $(s)$ corresponds to an example of formula with a simpler expression than the one proposed by \citet{Lantoine2011}. L2D labeled solutions were explicitely given by \citet{Lantoine2011} only in the 2D case, whereas L3D labeled solutions were also given by \citet{Lantoine2011} in the 3D case. The function sign is noted sg.}
\label{summary}
\end{figure*}

The $U$-motion is provided by the equation \ref{umotion}. The $W$-motion is provided by the equations \ref{wmotion1}, \ref{wmotion2} or \ref{wmotion3}. The $\phi$-motion is provided by $P_{\phi} \times [\ref{phiu}+(\ref{phiw1} \text{ or } \ref{phiw2} \text{ or }\ref{phiw3})]$. The time equation is provided by $[\ref{tu}+(\ref{tw1} \text{ or } \ref{tw2} \text{ or }\ref{tw3})]$. All the expressions are functions of $\mathcal{T}_0$ that is not the real time. We thus have implicit expressions as a function of time. The function $\tau(\mathcal{T}_0)$ is bijective but cannot be inversed  analytically, a numerical inversion is needed to derive the real time.

Besides, our 3D solutions can be easily applied to the 2D case. Indeed, in the 2D case, $P_\phi=0$ and thus, one of the roots for each polynomial $P_3$ and $Q_3$ is null: it could be $U_0$ or $U_{-}$ for $P_3$ (if $U_{+}=0$, there is no possible motion) and any roots of $Q_3$. We precise that in this case the $\phi$-motion is not important because the motion is planar. Compared with \cite{Lantoine2011}, our formulations are first developed for the 3D case and can be used easily for the 2D case, whereas \cite{Lantoine2011} gave only the methodology to obtain the 3D solutions based on 2D ones but not the expressions, which apparently leads to complex expressions.
\citet{Biscani2014} provided also the exact formulas for bounded and unbounded trajectories using the Weierstrass functions but this formulation is also difficult particularly because of the need to use the Inverse Weierstrass function, not implemented in all computer softwares and the need to work with complex values (e.g. the complex logarithm function). In this paper, we solved the motion for the bounded and unbounded trajectories in the 3D case; we provide the exact formulas for all cases, as well as the definitions used in \ref{appendixA}. We also highlight in table \ref{summary} which solutions are simpler compared with \citet{Lantoine2011}, which are completely new and which were only provided in the 2D case by \citet{Lantoine2011}.

Moreover, beyond the new exact solutions given in this paper, the derivation of our solutions based on Jacobi elliptic functions allows a good computing time and accuracy. \citet{Hatten2014} compared three types of solutions for the Stark effect: two exact ones, proposed by \citet{Lantoine2011} (Jacobi elliptic functions) and \citet{Biscani2014} (Weierstrass elliptic functions), and a numerical one by \citet{Pellegrini2014} (based on Taylor series). They compared the CPU time, the number of calls for each analytic elliptic function and the accuracy between \citet{Biscani2014} and \citet{Lantoine2011}. Even if we do not agree with the number of evaluations of each Jacobi elliptic function mentioned by \citet{Hatten2014} (e.g. to call $\cn$ and $\sn$ is similar to call $\am$ then $\cos$ and $\sin$: $\cn=\cos\circ\ \am$ and $\sn=\sin\circ\ \am$), two arguments show our solutions are efficient in terms of CPU time and accuracy : first, the solutions expressed in terms of Jacobi elliptic functions (such as in this paper or by \citet{Lantoine2011}) are more efficient than Weierstrass elliptic functions (used by \citet{Biscani2014}) ; second, several solutions given above are less complex to implement than those by \citet{Lantoine2011}, e.g. equation \ref{wmotion3}. Also, the analytical formulations are preferable for long duration motions.

\section{Conclusions}
We determined analytically the trajectories of the particles or spacecraft under the influence of both planetary gravity and stellar radiation pressure. We thus provide for the first time the complete exact solutions of the well-known Stark effect (effect of a constant electric field on the atomic Hydrogen's electron) with Jacobi elliptic functions, for both bounded and unbounded orbits. These expressions may be implemented for modeling spacecraft or particles trajectories: instead of solving the equation of the motion, based on differential equations, with numerical methods such as the Runge-Kutta method where one cumulates errors along the time, it is here possible to obtain precise expressions of the motion with only periodic errors, due to the precision on the evaluation of the elliptic functions used. In particular, we provide the analytical conditions for stable circular orbits. Moreover, we discuss about the possible issues inherent to the formalism used and the importance of being extremely careful with the routines implemented.

The formalism used here will allow us in a next paper to generalize the work by \citet{Bishop1989} to derive the exact neutral densities and escape flux in planetary exospheres, under the influence of both gravity and stellar radiation pressure. This is important for understanding the atmospheric structure and escape of planets in the inner solar system, as well as the atmospheric erosion during the early ages where the radiation pressure (and UV flux) of the Sun was extreme.

\section*{Acknowledgments}

This work was supported by the Centre National d'Études Spatiales (CNES).

\appendix

\section{Elliptic integrals}\label{appendixA}

In this paper, we use the three incomplete elliptic integrals  $F$, $E$ and $\Pi$:
\begin{equation}
\begin{array}{l}
\displaystyle F(\phi,k)=\int_{0}^{\phi} \! \dfrac{1}{\sqrt{1-k^2\sin^2\theta}}\,\mathrm{d}\theta\\\\
\displaystyle E(\phi,k)=\int_{0}^{\phi} \! \sqrt{1-k^2\sin^2\theta}\,\mathrm{d}\theta\\\\
\displaystyle \Pi(n;\phi,k)=\int_{0}^{\phi} \! \dfrac{1}{(1-n\sin^2\theta)\sqrt{1-k^2\sin^2\theta}}\,\mathrm{d}\theta
\end{array}
\label{goodformule}
\end{equation}

Sometimes, other formulas (shown below) are proposed with the change $\sin\theta=t$ but one needs to be very careful: this change is bijective only for $\theta\in [-\pi/2;\pi/2]$

\begin{equation}
\begin{array}{l}
\displaystyle \int_{0}^{x=\sin\phi} \! \dfrac{\sqrt{1-k^2t^2}}{\sqrt{1-t^2}}\,\mathrm{d}t
\displaystyle= \int_{0}^{\phi} \! \dfrac{\cos\theta}{|\cos\theta|}\sqrt{1-k^2\sin^2\theta}\,\mathrm{d}\theta\\\\
\displaystyle \int_{0}^{x=\sin\phi} \! \dfrac{1}{\sqrt{1-k^2t^2}\sqrt{1-t^2}}\,\mathrm{d}t
\displaystyle= \int_{0}^{\phi} \! \dfrac{\cos\theta}{|\cos\theta|}\dfrac{1}{\sqrt{1-k^2\sin^2\theta}}\,\mathrm{d}\theta\\\\
\displaystyle \int_{0}^{x=\sin\phi} \! \dfrac{1}{(1-nt^2)\sqrt{1-k^2t^2}\sqrt{1-t^2}}\,\mathrm{d}t
\displaystyle= \int_{0}^{\phi} \! \dfrac{\cos\theta}{|\cos\theta|}\dfrac{1}{(1-n\sin^2\theta)\sqrt{1-k^2\sin^2\theta}}\,\mathrm{d}\theta
\end{array}
\label{badformule}
\end{equation} 
These expressions are not exactly $E$, $F$ and $K$. They agree with the previous formulas \ref{goodformule} in the range $\phi\in]-\pi/2;\pi/2[$. \citet{Lantoine2011} did not precise which formulations they used. According to theirs formulas and results, they used the left-hand side of the equations \ref{badformule}. This may be a problem for bounded trajectories: for $\phi=\am(\mathcal{T}_0)\Longrightarrow x=\sn(\mathcal{T}_0)$, the integrals \ref{badformule} are not continuous contrary to \ref{goodformule}. Depending on the computer software, the routines and the definitions used for these functions, the results can show some issues (e.g. no continuous motion).

\bibliographystyle{model2-names}

\end{document}